\documentclass[trackchanges]{aastex701}
\usepackage{booktabs}
\usepackage{amsmath}
\usepackage{siunitx}

\begin{document}

\title{{Revisiting Gaussian Process Reconstruction for Cosmological Inference:\\ The Generalised GP (Gen GP) Framework}}

\author[gname=IceSheet]{Ruchika}
\affiliation{Departamento de Física Fundamental and IUFFyM, Universidad de Salamanca, E-37008 Salamanca, Spain}

\affiliation{INFN, Sezione di Roma, Piazzale Aldo Moro 2, I-00185 Roma, Italy}
\affiliation{Dipartimento di Fisica, Università di Roma "La Sapienza", Piazzale Aldo Moro 2, I-00185 Roma, Italy}
\email[show]{ruchika.science@usal.es}

\author[gname=IceSheet]{Purba Mukherjee}
\affiliation{Centre for Theoretical Physics, Jamia Millia Islamia, New Delhi-110025, India}
\email{purba16@gmail.com}

\author[gname=IceSheet]{Arianna Favale}
\affiliation{Dipartimento di Fisica, Università di Roma Tor Vergata, and INFN, Sezione di Roma 2, via della Ricerca Scientifica, 1, 00133, Roma, Italy}
\affiliation{Departament de Física Quàntica i Astrofísica, Universitat de Barcelona, and Institut de Ciències del Cosmos, c. Martí i Franqués, 1, 08028 Barcelona, Catalonia, Spain}
\email{afavale@roma2.infn.it}

\correspondingauthor{Ruchika}

\begin{abstract}

We investigate uncertainties in the estimation of Hubble constant ($H_0$) arising from Gaussian Process (GP) reconstruction, demonstrating that  the choice of kernel introduces systematic variations comparable to those arising from different cosmological models. To address this limitation, we introduce the {\it Generalized Gaussian Process} (Gen GP) framework, in which the Matérn smoothness parameter $\nu$ is treated as a free parameter, allowing for data-driven kernel optimization. Using the cosmic chronometer Hubble data, we find that while standard GP with $\Lambda$CDM mean function exhibits noticeable reconstruction differences between optimized and marginalized approaches, particularly at $z > 1$, Gen GP maintains  methodological consistency. 

In Gen GP, slight increases in $\chi^2$ per degree of freedom relative to standard GP, for both the zero-mean and $\Lambda$CDM prior mean cases, reflect added flexibility rather than performance degradation. Our results emphasize that robust cosmological inference requires treating kernel parameters as free variables and implementing full Bayesian marginalization to avoid artificial precision from fixed hyperparameters. As machine learning becomes central to cosmological discovery, the Gen GP framework provides a principled approach to model-independent inference that properly accounts for methodological uncertainties while maintaining necessary flexibility for reliable parameter estimation.

\end{abstract}


\section{Introduction}

Machine Learning has made a remarkable job in interpreting cosmological data in recent years \cite{Dvorkin, Ravanbakhsh:2017bbi}. As we anticipate greater opportunities in the next decade, challenges also arise in adopting Machine Learning (hereafter ML) methodologies and extracting meaningful insights \cite{Villaescusa-Navarro:2020rxg}. Its transformative potential lies in its ability to address cosmological issues such as identifying data clusters, handling sparse data and outliers, classifying datasets, and unraveling the physical intricacies within extensive datasets \cite{baron2019machine,  Goodfellow-et-al-2016}. The versatility of ML holds promise for revolutionizing the field of cosmology, but a comprehensive understanding of challenges is crucial to harness its full potential for groundbreaking research in the coming years \cite{ball2010data, Charnock:2020yur}.

The applications of ML in cosmology are diverse and impactful. Notably, ML has demonstrated its ability to significantly reduce scatter in mass estimates of clusters compared to traditional methods \cite{Ntampaka:2014ypa, Modi:2018cfi}. In the realm of weak lensing maps, where non-Gaussianities pose challenges in measurement and parameterization, ML emerges as a powerful solution \cite{wilde2022detecting, Wang, Schmelzle, Matilla}. Studies have demonstrated that these non-Gaussianities encode valuable cosmological information, leading to a remarkable tightening of parameter constraints by a factor of five or more \cite{Ribli:2019wtw, 2019MNRAS.487..104M}. Additionally, for next-generation Cosmic Microwave Background (CMB) experiments, ML is shown to provide competitive methods for extracting a high signal-to-noise ratio of the projected gravitational potential from CMB sky maps \cite{Caldeira:2018ojb, Wang, Munchmeyer:2019kng}. ML's reach extends to predicting structure formation \cite{Lucie-Smith:2020ris, HarnoisDeraps:2019xih} and classifying the earliest luminous sources that drove the reionization epoch \cite{hassan2020identifying}. Furthermore, in photometric surveys like the Vera Rubin C. Observatory (LSST) \cite{LSST:2018jkl, LSST:2008ijt}, ML is leveraged to classify supernovae, and this classification is subsequently utilized to obtain cosmological constraints \cite{Lochner:2016hbn}. The myriad applications underscore ML's pivotal role in advancing our understanding of cosmological phenomena and shaping the future of observational cosmology.

While ML cannot replace statistical reasoning for well-modeled phenomena, the increasing precision in cosmological data enhances the theoretical gaps in the current understanding of the universe \cite{Raveri:2018wln, Sellentin:2015waz}. The standard model of cosmology, a spatially flat $\Lambda$-cold-dark-matter ($\Lambda$CDM) framework with nearly scale-invariant primordial power spectra, provides the best description of cosmological observations, including CMB anisotropies, large-scale structure (LSS) clustering, the magnitude–redshift relation of type Ia supernovae (SNIa), and weak lensing of both the CMB and LSS (cosmic shear) \cite{planck2020, DES:2021wwk, Heymans:2020gsg}. However, several observational hints raise fundamental questions, the most pronounced being the Hubble tension -- now exceeding the 5$\sigma$ confidence level -- between the Hubble constant $H_0$ inferred from high-redshift CMB measurements ($H_0 = 67.36 \pm 0.54$ km/s/Mpc) \cite{planck2020} and that obtained from local SNIa observations ($H_0 = 73.04 \pm 1.04$ km/s/Mpc) \cite{Riess:2021jrx, Verde:2019ivm, di2021snowmass2021}.

 The existing state-of-the-art models (featuring both early and late time modifications to $\Lambda$CDM) struggle to explain all the cosmological datasets together \cite{Knox:2019rjx, Mortsell:2018mfj, Smith:2020rxx,Dutta:2018vmq,Ruchika:2020avj,Lonappan:2017lzt,Evslin:2017qdn,Patil:2023rqy,Forconi:2023izg,Forconi:2023hsj,Krishnan:2020obg}, paving the way for ML to play a vital role in the future of data-driven cosmological discovery. ML, however, poses a dual challenge: it accelerates discoveries in the face of upcoming data-intensive surveys \cite{Euclid:2019clj, LSST:2018jkl}, yet tempts researchers to prioritize expediency over deep understanding, particularly in the presence of complex or rapidly increasing data volumes \cite{Dvorkin, Jeffrey:2019fag}. Acknowledging that small biases in ML are currently overshadowed by other systematics, it is imperative for the future to meticulously address and account for biases to ensure the robustness and accuracy of ML applications in the evolving landscape of cosmological research \cite{Ntampaka:2019udw, Charnock:2020mdk, Villaescusa-Navarro:2020rxg}.

Artificial Neural Networks \cite{Geron2019, Goodfellow-et-al-2016}, Convolutional Neural Networks \cite{lecun2015deep}, Gradient Boosting Regression  \cite{Friedman2001}, Extra-Trees \cite{Geurts2006}, Support Vector Machines \cite{CortesVapnik1995, Murphy:2012}, Genetic Algorithms \cite{Holland1992}, and Gaussian Processes (GP) \cite{2006gpml.book.....R, Bishop2006} have emerged as powerful tools for deciphering intricate patterns within data, with GP notably finding applications in cosmological fields like parameter estimation \cite{Seikel:2012uu,Haridasu:2018gqm,Handley2019, Raveri:2018wln,Yang:2020bpv, Velazquez:2024aya, Mukherjee:2024cfq}, calibration of the distance ladders \cite{Gomez-Valent:2021hda,Liang:2022smf,Favale:2023lnp, li_testing_2023,Favale:2024lgp, Mukherjee:2024akt, Mukherjee:2024wix}, constraining CMB temperature evolution \cite{Ruchika:2025sbb}, measuring 
the sound horizon at the drag epoch \cite{Lemos:2023qoy},
and measuring Hubble constant \cite{shafieloo2012gaussian, Busti:2014dua,Gomez-Valent:2018hwc, Yang:2022jkf}. 

In this work, we provide a detailed study of the Gaussian Process framework, incorporating the Cosmic Chronometers (CC hereafter) data (see, e.g., \cite{Moresco:2022phi, Moresco:2024wmr} for dedicated reviews).
We introduced the Generalized Gaussian Process (Gen GP), an extension of traditional GP, that eliminates the inherent bias of the kernel. The results strongly emphasize the potential for continued progress in such cosmology-independent statistical techniques \cite{Ravanbakhsh:2017bbi, Berger:2018cub,Capozziello:2018jya}, reinforcing the assertion that data-driven inference frameworks are not only a promising future path but represent a dynamic field that requires ongoing refinement of methodologies to mitigate biases and ensure accurate training of cosmological datasets \cite{Sellentin:2015waz}. We find that both the shapes of the Hubble function $H(z)$ and its normalized definition, $E(z)=H(z)/H_0$, can be reconstructed with high accuracy when we keep the GP unbiased by choosing any particular kernel and cosmology-dependent prior mean function. While the impact of kernel and mean function choices has been discussed in previous literature (see, e.g. \cite{shafieloo2012gaussian, holsclaw2011nonparametric, prl.105.241302, Seikel:2012uu,Favale:2023lnp, Hwang:2022hla, Johnson:2025blf, Jiang:2025ilh, Mukherjee:2025ytj, Mukherjee:2024ryz, Mukherjee:2020ytg}), here we take a deeper step by introducing a modification of the conventional GP framework. This allows us not only to assess these dependencies more rigorously but also to enhance the robustness of the reconstructed functions. Furthermore, we estimate cosmological parameters using the reconstructed $H(z)$ and find that the results are consistent with those obtained directly from observational data.

This manuscript is organised as follows. Section \ref{sec:description} establishes the theoretical and methodological foundations essential for cosmological data interpretation through a dual approach. We first present the cosmological model landscape, examining various theoretical frameworks: the standard $\Lambda$CDM model, the constant equation of state $w$CDM, the dynamical Chevallier-Polarski-Linder (CPL) parametrization, and the cosmology model-independent Pad\'e approximation. We then develop the statistical inference framework using Gaussian Processes, detailing the different kernel choices and prior mean functions, culminating in the introduction of our Generalized GP approach. This comprehensive treatment allows systematic analysis of how both theoretical model assumptions and statistical methodology choices fundamentally shape cosmological parameter inference. Section \ref{sec:chisqana} presents the cosmological datasets and $\chi^2$ minimization techniques employed in our analysis. Gaussian Process training method and the Prediction Algorithm is described in section \ref{sec:gp_framework}. Section \ref{sec:analysisandres} present our results, discuss the implications for cosmological inference, and provide strong motivation for the Gen GP framework. Finally, Section \ref{sec:res} summarizes our conclusions and their broader implications for robust cosmological parameter estimation.

\section{The Dual Nature Inference: Cosmological Models and Statistical Methods}\label{sec:description}

\subsection{Cosmological Framework}

The flat $\Lambda$CDM model predicts the evolution of the normalized Hubble parameter with redshift as \cite{planck2020, Planck:2015fie}: 
\begin{equation}
\frac{H(z)}{H_{0}} = \left[\Omega_{m0}(1+z)^3 + (1-\Omega_{m0})\right]^{\frac{1}{2}},
\end{equation}
where $\Omega_{m0}$ represents the present-day matter density parameter and $H_0$ is the Hubble constant.

The $w$CDM model introduces a constant equation of state parameter $w$ for dark energy \cite{Weinberg:1988cp, Padmanabhan:2002ji}, modifying the normalized Hubble parameter evolution to:
\begin{equation}
     \frac{H(z)}{H_{0}} = \left[\Omega_{m0}(1+z)^3 + (1-\Omega_{m0})(1+z)^{3(1+w)}\right]^{\frac{1}{2}}.
\end{equation}

For enhanced flexibility in describing dark energy evolution, we employ the Chevallier-Polarski-Linder (CPL) parametrization \cite{Chevallier:2000qy, Linder:2002et}. The CPL model incorporates a redshift-dependent equation of state parameter $w(z) = w_{0} + w_{a}(1-a)=w_{0} + w_{a}\frac{z}{1+z}$, where $w_0$ and $w_a$ characterize the present-day equation of state and its evolution, respectively. The corresponding normalized Hubble parameter is:
\begin{equation}
     \frac{H(z)}{H_{0}} = \left[\Omega_{m0}(1+z)^3 + \left(1-\Omega_{m0}\right)f(z)\right]^{\frac{1}{2}},
\end{equation}
where $f(z)=\exp\left[3\int_0^z\frac{1+w(x)}{1+x}{\rm d} x\right]$ \cite{adachi2012analytical}.

We also investigate a Pad\'e approximation approach for dark energy modeling \cite{Aviles:2014rma}, which avoids specific parametric assumptions about the dark energy equation of state. This method models deviations from a cosmological constant through a Taylor expansion of the dark energy density around the present epoch ($z=0$):
\begin{equation}
    \rho_{de}(a) = \rho_{0} + \rho_{1} (1-a) + \rho_{2} (1-a)^2 + \ldots
\end{equation}
where $\rho_{0}$, $\rho_{1}$, and $\rho_{2}$ represent the present-day dark energy density and its first and second derivatives, respectively. Converting to redshift coordinates and applying a (2,2) Pad\'e approximation \cite{Gruber:2013wua}, the normalized Hubble parameter becomes:
\begin{equation}
     \frac{H(z)}{H_{0}} = \left[\Omega_{m0}(1+z)^3 + (1-\Omega_{m0}){\cal P}(z)\right]^{\frac{1}{2}},
\end{equation}
where
\begin{equation}
    {\cal P}(z)=\frac{1+P_{1}z+P_{2}z^2}{1+Q_{1}z+Q_{2}z^2},
\end{equation}
with $P_{1}$, $P_{2}$, $Q_{1}$, and $Q_{2}$ serving as the four expansion parameters. This Pad\'e parametrization encompasses a broad range of dark energy behaviors, including phantom-crossing scenarios and negative effective dark energy densities, while naturally reducing to $\Lambda$CDM in the limits of very low and very high redshifts.

\subsection{Statistical Framework: Gaussian Processes}

Gaussian Processes provide a non-parametric Bayesian approach for function reconstruction that has found extensive application in cosmological inference \cite{2006gpml.book.....R}. In cosmology, GP methods have been successfully employed to reconstruct the cosmic expansion history, matter distribution evolution, and structure growth from diverse observational datasets including SNIa, baryon acoustic oscillations (BAO), growth measurements, and CC (see, e.g, \cite{Seikel:2012uu, shafieloo2012gaussian, holsclaw2011nonparametric, Haridasu:2018gqm, Mukherjee:2020ytg, Mukherjee:2024ryz, Favale:2023lnp}).

The GP reconstruction framework depends critically on two fundamental components: the choice of kernel (covariance) function and the prior mean function \cite{MacKay2002}. These elements encode assumptions about the underlying physical process and substantially influence both the behavior and predictive capability of the GP reconstruction. Proper selection of these functions allows incorporation of physical knowledge while maintaining the flexibility necessary for reliable cosmological inference.

Given $n$ observational data points $y = (y_1,\ldots,y_n)$ at redshifts $z = (z_1,\ldots,z_n)$ with covariance matrix $C$, the objective is to reconstruct the underlying function $f^{*} = \left(f(z^{*}_1), \ldots, f(z^{*}_N)\right)$ at new points $z^{*} = (z^{*}_1,\ldots,z^{*}_N)$, typically with $N > n$.

A Gaussian Process represents a generalization of the multivariate Gaussian distribution to function space, completely specified by a mean function $\mu(z)$ and covariance function $k(z,z')$ \cite{Neal:1996}:
\begin{equation}
    f \sim \mathcal{N}\left(\mu(z),\, k(z,z')\right) 
\end{equation}
where $k_{ij} = k(z_i,z_j)$ is defines the covariance matrix elements of covariance $k(z,z')$ .

\subsubsection{Kernel Selection}\label{sec:kernelsec}

The most commonly employed covariance function in cosmological applications is the squared exponential (SE) kernel \cite{holsclaw2011nonparametric}:
\begin{equation}\label{eq:SE}
    k_{\mathrm{SE}}(r) = \sigma_f^2 \exp\left(-\frac{r^2}{l_f^2}\right)
\end{equation}
where $r = |z-z'|$, and the hyperparameters $\sigma_f$ and $l_f$ control the signal variance and characteristic length scale, respectively \cite{Seikel:2012uu}.

Additionally, we consider the Double Squared Exponential (DSE) kernel, constructed as the sum of two SE kernels:
\begin{equation}
k_{\mathrm{DSE}}(r) = \sigma_{f1}^2 \exp\left(-\frac{r^2}{l_{f1}^2}\right) + \sigma_{f2}^2 \exp\left(-\frac{r^2}{l_{f2}^2}\right),
\end{equation}
which increases modeling flexibility at the cost of doubling the number of hyperparameters.

For greater generality, one employs the Mat\'ern class of covariance functions \cite{Seikel:2012uu}, which provides a more flexible framework for modeling correlation structures. The general Mat\'ern kernel is defined as \cite{Rasmussen:2010}: 
\begin{equation}\label{eq:gengp}
    k_\nu(r) = \sigma_f^2 \frac{2^{1-\nu}}{\Gamma(\nu)} \left( \frac{\sqrt{2\nu}r}{l_f} \right)^\nu K_\nu \left( \frac{\sqrt{2\nu}r}{l_f} \right) \, ,
\end{equation}
where $\Gamma$ is the gamma function, $K_\nu$ is the modified Bessel function of the second kind \cite{Abramowitz:1964}, and $\nu$ is a positive parameter controlling the smoothness of the process.

For half-integer values of $\nu$, the Matérn kernel simplifies to closed-form expressions \cite{holsclaw2011nonparametric, Seikel:2012uu}, viz.
\begin{eqnarray}\label{eq:kernel}
    k_{\nu=3/2}(r) &= \sigma_f^2 \left(1+ \frac{\sqrt{3}r}{l_f}\right) \exp\left(-\frac{\sqrt{3}r}{l_f}\right), \\ \nonumber
    k_{\nu=5/2}(r) &= \sigma_f^2 \left(1+ \frac{\sqrt{5}r}{l_f} + \frac{5r^2}{3l_f^2}\right) \exp\left(-\frac{\sqrt{5}r}{l_f}\right), \\ \nonumber
    \quad \text{and so on}.
\end{eqnarray}
The parameter $\nu$ determines the differentiability properties of the reconstructed function: processes with $\nu = 1/2$ are continuous but not differentiable, while larger values of $\nu$ correspond to increasingly smooth functions. In the limit $\nu \to \infty$, the Mat\'ern kernel converges to the squared exponential form \cite{Neal:1996}. Previous studies suggest that $\nu \leq 7/2$ provides adequate flexibility for most cosmological applications \cite{2006gpml.book.....R, Favale:2023lnp, Banerjee:2023evd, Mukherjee:2024ryz}.

For enhanced generality and to avoid kernel selection bias, we employ the 
\textbf{Generalized Gaussian Process (Gen GP) framework}, which utilizes 
the Matérn class of covariance functions (see Equation \ref{eq:gengp}). Unlike 
standard GP implementations that fix the smoothness parameter $\nu$ to 
specific values, Gen GP treats $\nu$ as a free parameter (with uniform prior 
$\log_{10}(\nu) \in \mathcal{U}[-2,1]$ as detailed in Table \ref{table:basic_prior}), allowing the 
data to determine the optimal smoothness properties.

\subsubsection{Prior Mean Function Selection}

The prior mean function $\mu(z)$ represents the initial assumption about the underlying function before observing data \cite{shafieloo2012gaussian}. This choice significantly impacts reconstruction quality, as an appropriate prior can reduce the dynamic range over which the GP must extrapolate from the data \cite{holsclaw2011nonparametric}.

The most agnostic choice is the zero mean function, $\mu(z) = 0$, which avoids imposing specific cosmological assumptions. However, this approach may require large values of $\sigma_f$ to accommodate the full range of observational data, potentially necessitating large correlation lengths $l_f$ that could smooth away important features and compromise derivative reconstructions \cite{Seikel:2012uu}.

Alternatively, one can adopt a model-based prior mean function, such as the $\Lambda$CDM prediction. In this case, the GP models deviations from the assumed cosmological model as
\begin{equation}
    \mu(z) = y_{\text{obs}} - y_{\Lambda\text{CDM}}(z) \, ,
\end{equation}
where $y_{\Lambda\text{CDM}}(z)$ represents the theoretical prediction. While this approach can improve reconstruction efficiency, it introduces the risk that final results retain memory of the initial assumption, potentially biasing cosmological parameter estimates \cite{Seikel:2012uu, Favale:2023lnp, Hwang:2022hla}.

In our analysis, we examine both, a zero mean function for a more agnostic approach, and a $\Lambda$CDM-based mean function to assess the impact of theoretical priors on parameter inference \cite{verde2010statistical}. This comparative approach establishes a baseline that allows us to quantify the systematic effects of complex mean function choice on cosmological conclusions \cite{shafieloo2012gaussian}.

\section{GP Methodology: Hyperparameter Sampling and Likelihood Evaluation}\label{sec:chisqana}

This section focuses on generalising GP reconstruction to infer the Hubble function $H(z)$ using CC measurements. These data have been collected over the past two decades via the differential age technique \cite{Jimenez:2001gg}, utilising massive and passively evolving galaxies observed up to redshift $z \lesssim 2$ from different galaxy surveys. 

Our analysis employs 32 cosmic chronometer measurements spanning redshifts 
$0.07 \leq z \leq 1.965$, compiled from multiple surveys mentioned in \cite{Moresco:2022phi, Moresco:2024wmr}.
These measurements, obtained via 
the differential age method applied to massive, passively evolving galaxies, 
provide direct constraints on $H(z)$ independent of distance ladder 
calibrations or cosmological model assumptions. The full dataset is also available at Gitlab\footnote{\url{https://gitlab.com/mmoresco/CC_covariance}}.

\subsection{Dataset Choice and Methodological Focus}\label{sec:datasetchoices}
Although a complete analysis would benefit from including multiple observational probes such as SNIa, BAO and CMB data, our work remains primarily methodological in nature. We aim to advance the GP framework for cosmological inference, addressing key issues such as hyperparameter uncertainties and kernel selection bias.

The deliberate choice to concentrate on CC data allows us to isolate and study the systematic effects inherent in GP reconstruction methods without the additional complexity introduced by combining heterogeneous datasets, each with its own systematic uncertainties and selection effects \cite{Sun:2021pbu}. This allows us to isolate the methodological performance of different GP implementations in parameter inference and uncertainty quantification. 

The extension to multi-probe cosmological analyses, while scientifically valuable, is beyond the scope of this work and represents a natural direction for future research. Once the methodological foundations established here are properly validated and understood, subsequent studies can apply these improved GP techniques to comprehensive datasets to obtain robust cosmological constraints. We plan to pursue such multi-dataset analyses in our future work, building upon the methodological framework developed in this paper.

\subsection{Parameter Inference and Sampling Strategy}

The core challenge in our approach is the efficient sampling of the GP hyperparameters, specifically $\sigma_f$, $l_f$, and, in the Gen GP case, an additional parameter $\nu$. For restricted GP models using fixed values of $\nu$ (e.g., Matérn kernels with $\nu = 3/2$, $5/2$, $7/2$, $9/2$, or SE and DSE), the hyperparameter space consists of $\{\sigma_f, l_f\}$ in addition to the standard cosmological parameters $\Theta^*$. In the Gen GP framework, $\nu$ is treated as a free parameter, increasing the dimensionality of the parameter space to $\Theta = \{\sigma_f, l_f, \nu\}$.

Sampling these parameters in their natural units (linear-space) is inefficient due to their wide dynamic ranges and often complex posterior shapes. To overcome this, we adopt a logarithmic sampling approach, assigning flat priors in log-space, namely $\log_{10}(\sigma_f)$, $\log_{10}(l_f)$,  and $\log_{10}(\nu)$.

This transformation allows for more uniform exploration of parameter space and improves convergence. For $\sigma_f$ and $l_f$, which can range from $10^{-5}$ to $10^{+5}$, log-sampling ensures efficient coverage. For $\nu$, which typically ranges from $10^{-2}$ (non-differentiable) to values above 1 (smooth functions), the log-scale avoids inefficient sampling near zero and improves numerical stability during matrix inversion.

Throughout the training process, sampling is performed entirely in log-space. When evaluating the likelihood and generating predictions, parameters are transformed back to their physical units: $\sigma_f = 10^{\log_{10}(\sigma_f)}$, $l_f = 10^{\log_{10}(l_f)}$, $\nu = 10^{\log_{10}(\nu)}$. This ensures the GP covariance matrix is constructed correctly while preserving the computational efficiency gained during sampling. 

The Matérn kernel, which relies on the additional $\nu$ to define smoothness, receives the back-transformed $\nu$ value at each step, allowing the data to determine the appropriate level of differentiability. Posterior samples are also transformed back for final inference and uncertainty estimates. 

Proper handling of these transformations is critical for avoiding biases and underestimation of uncertainties. The inclusion of $\nu$ as a free parameter in the Gen GP framework is particularly important, as it allows the data to inform the optimal smoothness, thereby reducing the risk of kernel selection bias.

\subsection{Bayesian Framework \& Likelihood Evaluation}
Parameter inference is carried out by maximising the GP log-marginal likelihood:
\begin{equation}\label{eq:Likelihood}
\ln \mathcal{L}(\Theta, \Theta^*) = -\frac{1}{2} \mathbf{y}^T K_y^{-1} \mathbf{y} - \frac{1}{2} \ln |K_y| - \frac{n}{2} \ln(2\pi),
\end{equation}
where $\mathbf{y} = \mathbf{H}_{\text{obs}} - \boldsymbol{\mu}(\mathbf{z})$ represents the residual vector between observations and the GP prior mean, and $K_y = K(\mathbf{X}, \mathbf{X}^*) + C$ is the total covariance matrix, combining the kernel matrix $K$ and the observational noise matrix $C$.

The first term in Eq.~\eqref{eq:Likelihood} quantifies the fit to the data, while the second term penalises overly complex models, functioning as an automatic Occam's razor. The final term ensures proper normalisation but does not affect parameter estimation directly.

To evaluate the inverse and determinant of $K_y$, we employ Cholesky decomposition, which guarantees numerical stability and computational efficiency across a wide range of hyperparameter values, including extreme or near-degenerate cases.

The hyperparameters sampled in log-space are converted back to physical units at each likelihood evaluation step for the proposed parameter set $\{\Theta, \Theta^*\}$ to ensure the GP covariance matrix is correctly constructed. This log-normal likelihood structure forms the basis of our Bayesian framework and naturally incorporates both goodness-of-fit and model complexity.

\subsubsection{Cosmological Model Analysis}
For comparison with GP-based approaches, we also perform standard MCMC 
analysis for parametric cosmological models ($\Lambda$CDM, $w$CDM, CPL, 
Padé) using the same CC dataset. The likelihood is:
\begin{equation}
\mathcal{L}(\Theta^*) = \exp\left(-\frac{1}{2}\chi^2\right), \quad 
\chi^2 = [H_{\text{obs}} - H_{\text{model}}(z;\Theta^*)]^T 
C^{-1} [H_{\text{obs}} - H_{\text{model}}(z;\Theta^*)]
\end{equation}
This provides model-dependent baselines for assessing GP reconstruction 
performance.

\subsection{Implementation Details}
We use the publicly available \texttt{emcee}\footnote{\url{https://emcee.readthedocs.io/en/stable/}} sampler \cite{Foreman_Mackey_2013} to generate Markov Chain Monte Carlo chains, and analyse the resulting samples with \texttt{GetDist}\footnote{\url{https://getdist.readthedocs.io/en/latest/}} \cite{Lewis:2019xzd}. Our analysis pipeline also utilizes \texttt{NumPy}\footnote{\url{https://numpy.org/}} \cite{numpy}, \texttt{SciPy}\footnote{\url{https://scipy.org/}} \cite{scipy}, \texttt{Astropy}\footnote{\url{http://www.astropy.org}} \cite{astropy1,astropy2,astropy3}, and \texttt{Matplotlib}\footnote{\url{https://matplotlib.org}} \cite{matplotlib}. For standard GP implementation, we use the \texttt{GaPP}\footnote{\url{https://github.com/carlosandrepaes/GaPP}} package \cite{Seikel:2012uu} as a baseline for our analysis.

We adopt uniform priors over wide ranges for all parameters (as shown {in Table \ref{table:basic_prior}), allowing the data to drive the inference without introducing strong prior assumptions. This is especially important for the GP hyperparameters, where restrictive priors can constrain the flexibility of the reconstruction and underestimate the associated uncertainties.

By allowing $\nu$ to vary freely in the Gen GP case and sampling all GP hyperparameters in log-space, our approach provides a more flexible and unbiased framework for non-parametric cosmological inference.

\begin{table*}[htb]\label{table:reskernel}
\centering
\resizebox{0.89\linewidth}{!}{%
\begin{tabular}{lSSSSSS}
\toprule
{Kernel Type} & {$H_{0}$ (km/s/Mpc)} & {$\sigma_{H_0}$ (km/s/Mpc)} & {$\sigma_{H_0}/H_0$ (\%)} & {$\sigma_f$} & {$l_f$} & $\chi^{2}/{\rm dof}$\\
\midrule
Matérn 3/2 & 69.35 & 6.34 & 9.13 & 140.01 & 4.29 &  0.42\\
Matérn 5/2 & 69.28 & 5.36 & 7.74 & 132.54 & 2.90 &  0.45\\
Matérn 7/2 & 68.87 & 5.08 & 7.37 & 133.20 & 2.61 & 0.47 \\
Matérn 9/2 & 68.50 & 4.95 & 7.23 & 135.57 & 2.54 &  0.48 \\
SE & 67.18 & 4.68 & 6.97 & 148.81 & 2.53 &  0.49\\
DSE & 67.33 & 6.39 & 9.49 & \multicolumn{1}{c}{[154.51, 10.08]} & \multicolumn{1}{c}{[3.19, 0.35]} & 0.43\\
\bottomrule
\end{tabular}}
\caption{Results at 68\% C.L. on $H_0$ with standard GP using different kernels and the values of their corresponding hyperparameters. Here, the a priori mean function is set to 0. The last column reports the values of $\chi^2/\mathrm{dof}$, where the degrees of freedom ({\rm dof}) are calculated as the number of data points minus fitted hyperparameters. See Sec. \ref{sec:compare} for details.}
\label{tab:Matérn_results}
\end{table*}

\begin{table*}[htbp]
\centering
\resizebox{0.6\linewidth}{!}{%
\begin{tabular}{lSSSS}
\toprule
Model & {$H_0$ (km/s/Mpc)} & {$\sigma_{H_0}$ (km/s/Mpc)} & {$\sigma_{H_0}/H_0$ (\%)} & $\chi^{2}/{\rm dof}$ \\
\midrule
$\Lambda$CDM & 68.84 & 4.11 & 5.97 & 0.48 \\
wCDM & 70.11 & 5.64 & 8.05 & 0.50\\
CPL & 70.97 & 6.04 & 8.51 & 0.51\\
\bottomrule
\end{tabular}}\caption{Comparison of $H_0$ values across different cosmological models. See Sec. \ref{sec:compare} for details.}

\label{tab:cosmological_models}
\end{table*}

\section{Gaussian Process Prediction Framework}\label{sec:gp_framework}

Given a set of observed data $\mathbf{y}$, our goal is to estimate the function values $\mathbf{f}^*$ at new test points $\mathbf{X}^*$ where we need to calculate the predictions. The conditional distribution of $\mathbf{f}^*$, given the observed data and training inputs $\mathbf{X}$, follows a multivariate Gaussian:
\begin{equation} \label{eq:predict}
\mathbf{f}^* \mid \mathbf{X}^*, \mathbf{X}, \mathbf{y} \sim \mathcal{N}\left(
\overline{\mathbf{f}^*},\, \mathrm{Cov}(\mathbf{f}^*)
\right) \;,
\end{equation}
with the predictive mean and covariance given by,
\begin{equation} \label{eq:mean}
\overline{\mathbf{f}^*} = {\mu}^* + K(\mathbf{X}^*, \mathbf{X}) \left[ K(\mathbf{X}, \mathbf{X}) + C \right]^{-1} (\mathbf{y} - {\mu})\,,
\end{equation}
and 
\begin{equation}
\mathrm{Cov}(\mathbf{f}^*) = K(\mathbf{X}^*, \mathbf{X}^*)
 - K(\mathbf{X}^*, \mathbf{X}) \left[ K(\mathbf{X}, \mathbf{X}) + C \right]^{-1} K(\mathbf{X}, \mathbf{X}^*)\,.
\end{equation}
Here, $K(\mathbf{X}, \mathbf{X})$ is the covariance matrix at the training inputs, $K(\mathbf{X}^*, \mathbf{X})$ denotes the cross-covariance between test and training points, $C$ accounts for the observational noise, and ${\mu}$ and ${\mu}^*$ are the prior mean functions evaluated at the training and test points, respectively. The predictive uncertainties at the test points are obtained from the diagonal of the covariance matrix $\mathrm{Cov}(\mathbf{f}^*)$.

In practice, we implement this prediction using the following \texttt{GaPP} code wrapper function \texttt{prediction}

which returns the predicted mean and standard deviation. This function numerically allows us to evaluate the expression in Eq.~\eqref{eq:mean}, using provided training data $z_{\text{CC}}$, test points $z \equiv z_{\text{pred}}$, observational values $H_{\text{CC}} \equiv H(z_\text{CC})$, data covariance $\Sigma_{\text{CC}}$, and the hyperparameters $\lbrace\Theta, \Theta^{*} \rbrace$.

Notably, both the choice of prior mean function and the treatment of hyperparameters influence the inferred results. The simplest strategy is to fix the GP hyperparameters and the parameters governing the mean function (e.g., $H_0$, $\Omega_m$ in case of a $\Lambda$CDM prior mean) to their best-fit values. This method may yield tighter constraints with smaller error bars, as it conditions the reconstruction on a single point in parameter space. However, it does not fully capture the underlying model uncertainty, since it ignores the posterior spread in both the kernel and mean function parameters. In our implementation, 
we account for the full posterior over both kernel and mean function parameters, resulting in more realistic reconstructions.

\section{Analysis and Interpretation of Results}\label{sec:analysisandres}

\subsection{Kernel vs Cosmology: A Comparative Study} \label{sec:compare}

To assess the robustness of standard GP reconstruction and its sensitivity to the choice of kernel, we examine how different covariance functions affect the inferred value of the Hubble constant $H_0$. Table \ref{tab:Matérn_results} presents the results obtained using various kernel types, which are described in Section \ref{sec:kernelsec}. 

Our findings indicate that the choice of kernel impacts both the central value and uncertainty of $H_0$. The inferred mean values span from 67.18 to 69.35 km/s/Mpc, although they are all compatible within 1$\sigma$. The Matérn kernels show a systematic trend: increasing the smoothness parameter $\nu$ (i.e., using smoother kernels) systematically lowers the predicted $H_0$ and reduces its associated uncertainty. For instance, the relative error decreases from 9.13\% for Matérn 3/2 to 7.23\% for Matérn 9/2.

The SE kernel predicts the lowest $H_0$ value (67.18 km/s/Mpc) with the smallest relative uncertainty (6.97 $\%$), reflecting its highly smooth nature. Despite introducing additional hyperparameters to capture multi-scale correlations, the DSE kernel yields results comparable to the lower-order Matérn kernels with higher uncertainties. The reduced chi-squared values remain consistently below 0.5, indicating acceptable fits despite variations in the smoothness assumptions.

When comparing different cosmological models (as shown in Table~\ref{tab:cosmological_models}), we observe similar trends. The $\Lambda$CDM model provides the most precise $H_0$ estimate (68.84 km/s/Mpc) with a relative uncertainty of 4.11\%. More flexible models, such as CPL, which introduce additional parameters, result in higher $H_0$ values, but with correspondingly larger uncertainties.  We also tested the Padé(2,2) parametrization \cite{Gruber:2013wua}, 
but found it poorly constrained by CC data alone due to parameter 
degeneracies. This is consistent with previous 
findings that Padé approximations require multi-probe datasets for 
stable constraints \cite{Capozziello:2018jya}.

{In summary, both the choice of kernel and the cosmological model in the GP framework significantly influence the inferred value of the Hubble constant and its precision. Smoother kernels generally produce tighter constraints, while increased model complexity tends to widen uncertainties despite potential shifts in central values.}

\begin{table*}[htb]\label{tab:prior}
\centering
\resizebox{0.95\linewidth}{!}{%
\begin{tabular}{lcc}
\toprule
\textbf{Parameter}  & \textbf{Priors} & \textbf{Description}\\ 
\midrule
  & \textbf{ Kernel Hyperparameters} & \\ 
\hline
$\log_{10}$($\sigma_f$) &  $\mathcal{U}[-5, 5]$ & overall amplitude of the correlation\\
$\log_{10}$($l_f$) &  $\mathcal{U}[-5, 5]$ &  coherence length of correlation \\
$\log_{10}$($\nu$) & $\mathcal{U}[-2,1]$ & defines the smoothness of the kernel\\
\hline
  & \textbf{Mean Function Parameters} & \\ 
\hline
$H_{0,\, \rm hyper}$ & $\mathcal{U}[50, 90]$ & Representative present expansion parameter in the mean function\\
$\Omega_{m0, \, \rm hyper}$ & $\mathcal{U}[0.1, 0.5]$ & Representative present matter density parameter in the mean function \\
\bottomrule 
\end{tabular} }
\caption{List of hyperparameters used in the training Gen GP and their corresponding priors. Uniform priors are chosen for all hyperparameters.}
\label{table:basic_prior}
\end{table*}

\begin{figure*}[htb]
    \centering
\includegraphics[width=\textwidth]{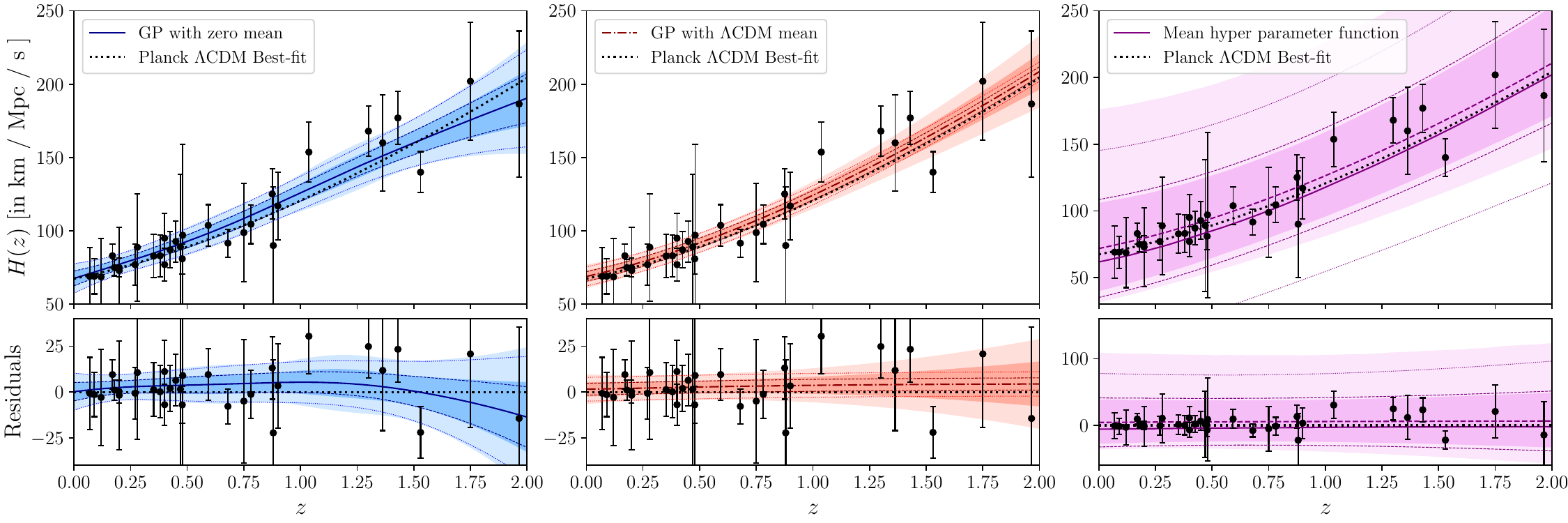}
\caption{Reconstructed evolution of the Hubble parameter $H(z)$ using GP with a Mat\'ern $\nu = 3/2$ kernel (top panel). The black dotted curve denotes the \textit{Planck} best-fit $\Lambda$CDM model. The blue and red shaded regions indicate the $2\sigma$ confidence intervals obtained using a zero-mean prior function and a $\Lambda$CDM mean prior function, respectively, using the full marginalized posteriors, while the dashed and dotted lines show the 1$\sigma$ and 2$\sigma$ confidence intervals when optimized hyperparameters are employed. The distribution of the mean function for the corresponding hyperparameters is shown in magenta. Bottom panel shows residuals with respect to the \textit{Planck} $\Lambda$CDM best-fit model. See Secs. \ref{sec:reconstruction_comparison} and \ref{sec:diss_mu} for details.}    \label{fig:gp}
\end{figure*}

\begin{figure*}[htb]
    \centering
\includegraphics[width=\textwidth]{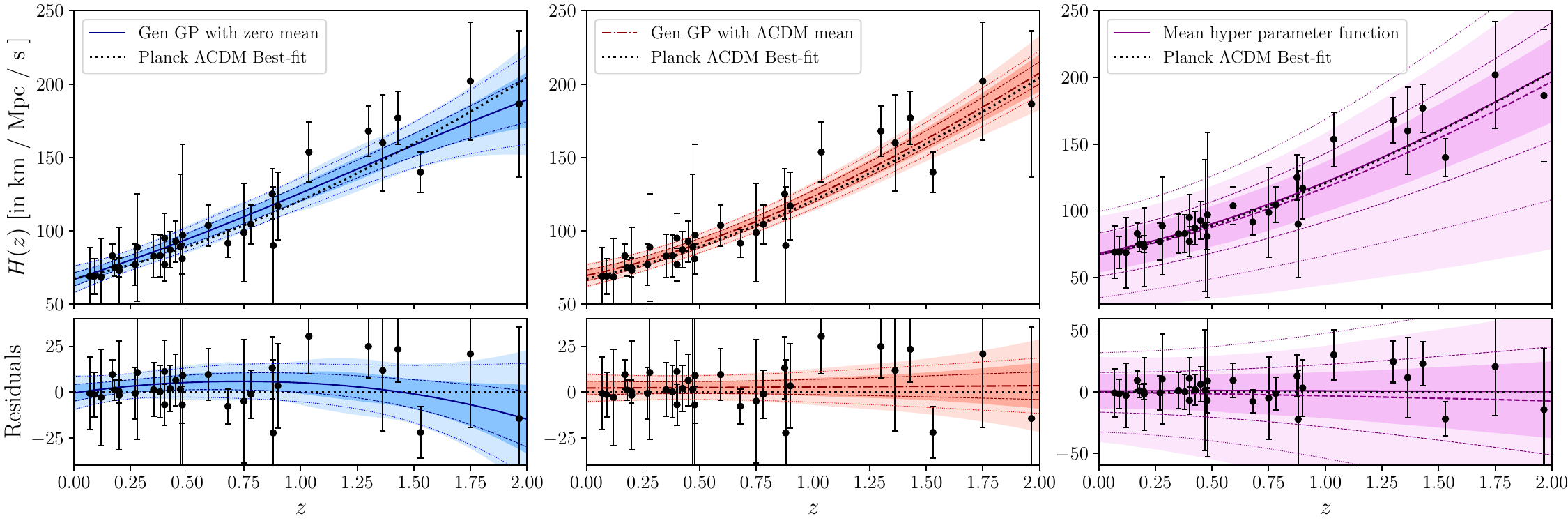}
    \caption{Reconstructed evolution of the Hubble parameter $H(z)$ using Gen GP (top panel). The black dotted curve denotes the \textit{Planck} best-fit $\Lambda$CDM model. The blue and red shaded regions indicate the $2\sigma$ confidence intervals obtained using a zero-mean prior function and a $\Lambda$CDM mean prior function, respectively, using the full marginalized posteriors, while the dashed and dotted lines show the 1$\sigma$ and 2$\sigma$ confidence intervals when optimized hyperparameters are employed. The distribution of the mean function for the corresponding hyperparameters is also shown in magenta. Bottom panel shows residuals with respect to the \textit{Planck} $\Lambda$CDM best-fit model. See Secs. \ref{sec:reconstruction_comparison} and \ref{sec:diss_mu} for details.}
    \label{fig:gengp}
\end{figure*}

\begin{table*}[htb]
\centering
\small
\resizebox{0.8\linewidth}{!}{%
\begin{tabular}{lcccccc}
\toprule
 Parameter & $\sigma_f$ & $l_f$ & $\nu_f$ & $H_{0,\, \rm hyper}$ & $\Omega_{m0, \, \rm hyper}$ \\
 \midrule 
 Zero mean &  $225.412^{+543.440}_{-117.234}$ &  $5.550^{+16.454}_{-3.427}$  & $2.864^{+3.825}_{-1.640}$ & - & - \\[1ex] 
$\Lambda$CDM mean & $33.26^{+91.93}_{-25.63}$   & $6.01^{+2.760}_{-3.070}$ & $3.96^{+2.043}_{-2.141}$ & $68.218^{+14.056}_{-14.485}$ & $0.310^{+0.110}_{-0.099}$  \\ \bottomrule 
\end{tabular}}
\caption{Constraints on hyperparameters with Gen GP when the a priori mean function is zero or $\Lambda$CDM. The Matérn Parameter $\nu$ is free, and the constraints error bars quoted are at $1\sigma$ confidence interval. See Sec. \ref{sec:hyperparameter_interpretation}. }\label{table:gengp}
\end{table*}

\subsection{Hyperparameter Treatment: Optimization vs Marginalization} \label{sec:reconstruction_comparison}

We compare two GP reconstruction strategies: utilizing the optimized hyperparameters (best-fit values) versus incorporating the full marginalized posteriors within the prediction function. In Figures \ref{fig:gp} and \ref{fig:gengp}, the shaded regions depict reconstructions using the full marginalized posteriors, while the dashed and dotted lines show the 1$\sigma$ and 2$\sigma$ confidence intervals when optimized hyperparameters are employed.

The leftmost panels of both figures reveal that there is no appreciable difference between these two reconstruction approaches, when employing a zero mean function\footnote{This is consistent with previous findings in \cite{Favale:2023lnp}, where the reconstructed shape of $H(z)$ obtained with standard GP has been found to differ by 5\% at most between marginalization and optimization procedure.}. This demonstrates that for zero-mean implementations, utilising best-fit hyperparameter values yields reconstructions equivalent in quality to those obtained from full posterior chains. This equivalence holds for both standard GP and Gen GP approaches with zero mean functions.

In contrast, the middle panel, which employs the $\Lambda$CDM mean function, reveals a notable divergence between the two reconstruction outcomes. The predicted $H(z)$ values obtained from marginalised posteriors differ significantly from those derived using the optimised framework, particularly at redshifts $z > 1$. This discrepancy suggests that the choice of reconstruction method and the treatment of hyperparameters becomes crucial when incorporating cosmological priors through the mean function.

A particularly striking observation is that this reconstruction discrepancy is substantially reduced in the generalised GP compared to the standard GP implementation. The standard GP exhibits considerably larger discrepancies between the two methods, while Gen GP demonstrates much better agreement between marginalised and optimised approaches. 

This enhanced consistency highlights Gen GP as the preferred choice for cosmological reconstructions when utilising the $\Lambda$CDM mean function, as it provides more robust results that are less sensitive to the specific implementation of hyperparameter treatment.

\begin{table*}[htb]
\centering
\small
\resizebox{0.8\linewidth}{!}{%
\begin{tabular}{lcccccc}
\toprule
 Parameter  & $H_{0,\, \rm hyper}$ & $H_0$(zero mean function)  & $H_0$($\Lambda$CDM mean function)  \\
 \midrule 
GP &  $61.647^{+44.294}_{-21.774}$ & $68.306^{+5.180}_{-5.177}$ & $68.914^{+4.064}_{-4.002}$ \\[1ex] 
Gen GP & $68.218^{+14.056}_{-14.485}$ & $68.101^{+4.975}_{-4.909}$  &  $69.563^{+4.150}_{-4.093}$ \\ \bottomrule 
\end{tabular}}
\caption{Constraints on hyperparameter $H_{0, \, \rm hyper}$ when the a priori mean function $\Lambda$CDM, vs predicted cosmological $H_0$ within Gen GP framework for both zero mean vs $\Lambda$CDM mean function scenarios. The Matérn kernel order parameter $\nu$ is free, and the constraints error bars quoted are at $1\sigma$ confidence interval. All values in km/s/Mpc. See Sec. \ref{sec:gengp_res}.} \label{Table:finalgpgengp}
\end{table*}

\begin{figure*}[htb]
    \centering
\includegraphics[width=0.49\textwidth]{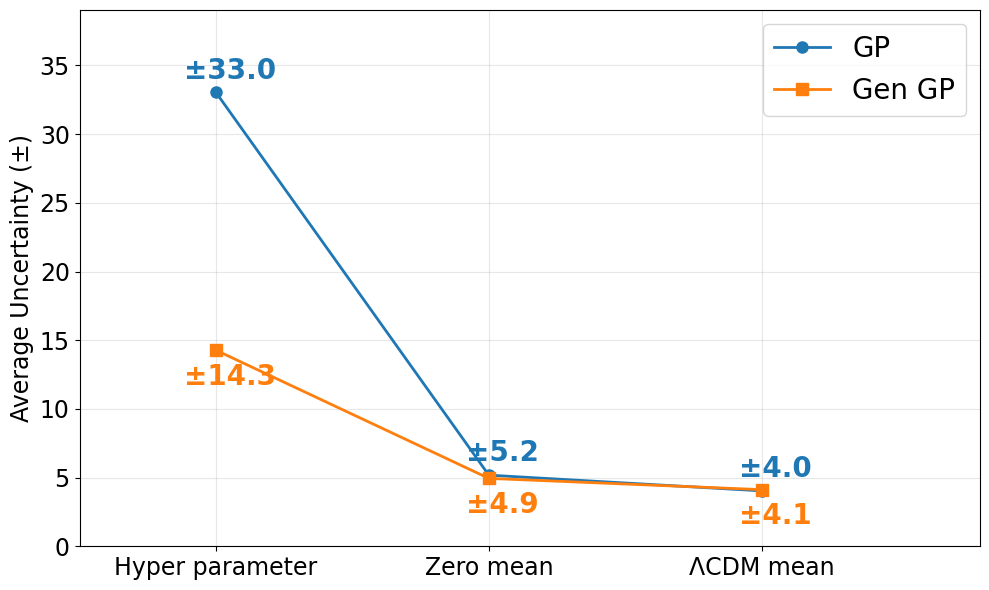} \hfill \includegraphics[width=0.49\textwidth]{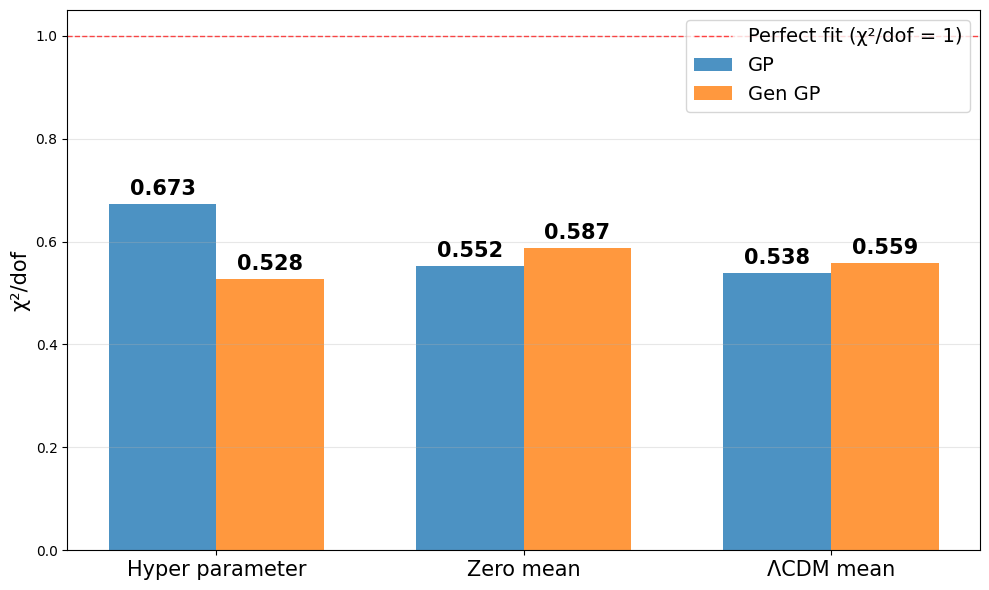}
    \caption{ The left panel shows the variation of uncertainty in the Hubble constant parameter for both GP and GenGP reconstructions, using both zero mean and $\Lambda$CDM mean functions. For comparison, we also include the case where only the mean function (assumed as $\Lambda$CDM) is used, without the full GP or GenGP modeling, prior to applying the covariance matrix (Eq. \ref{eq:meanfn}). The right panel presents a comparison of the corresponding $\chi^2$ values for each case.
}
    \label{fig:cc_var}
\end{figure*}

\subsection{Dissecting $\mu^{\star}$: Role of the GP mean function}\label{sec:diss_mu}

The rightmost panels in Figures \ref{fig:gp} and \ref{fig:gengp} provides crucial insight into the impact of the mean function $\mu^{\star}$ in cosmological reconstruction. There, we plot $H(z)$ using only the mean function component,
\begin{equation}\label{eq:meanfn}
    H(z)_{\mu^{\star}} = H_{0, \, \rm hyper} \left[ \Omega_{m0, \, \rm hyper} (1+z)^3 + \right. 
 \hfill \left.(1- \Omega_{m0, \, \rm hyper}) \right]^{\frac{1}{2}}\,.    
\end{equation}
This expression corresponds to $\mu^{\star}$ in Equation \eqref{eq:mean} and represents the cosmological prior embedded within the full $H(z)$ reconstruction shown in the middle panels. By isolating this component, we can examine how the hyperparameter $H_{0, \, \rm hyper}$ differs from the cosmologically inferred $H_0$.

Our MCMC analysis yields $H_{0,\, \rm hyper}=61.647^{+44.294}_{-21.774}$ km/s/Mpc for standard GP and $H_{0,\, \rm hyper}=68.218^{+14.056}_{-14.485}$ km/s/Mpc for Gen GP. These large uncertainties reflect the broad prior range and inherent flexibility built into the hyperparameter framework. However, the final reconstructed Hubble function exhibits significantly tighter constraints, with uncertainties of approximately $\pm 5$ km/s/Mpc at $z = 0$, as evident in the leftmost and middle panels of both figures. Similar findings are summarised in Table \ref{Table:finalgpgengp}.

This apparent contradiction stems from the fundamental distinction between the hyperparameter $H_{0,\, \rm hyper}$ and the actual reconstructed $H(z)$ at $z = 0$. The hyperparameter serves as a regularisation framework that enables the Gaussian process to capture the general functional form of cosmological expansion, while the final reconstruction represents a data-driven synthesis incorporating information from the mean function, kernel correlations, and observational constraints.

Our analysis employs both maximum a posteriori (MAP) estimates of hyperparameters and marginalisation over their full posterior distributions. Notably, we demonstrate that when utilising the $\Lambda$CDM mean function, the full marginalization approach provides enhanced reconstruction performance compared to MAP treatment, particularly in terms of consistency between different GP implementations.

\subsection{Interpreting GP Hyperparameters vs Physical Cosmological Parameters}\label{sec:hyperparameter_interpretation}

The GP frameworks employed in our analysis includes hyperparameters $H_{0,\, \rm hyper}$ and $\Omega_{m0, \, \rm hyper}$ within the mean function. 
While these two representative parameters are cosmologically motivated, they serve a fundamentally different role than their physical counterparts in cosmological models.

In traditional cosmological analysis, $H_0$ refers to the Hubble constant, physically interpreted as the current expansion rate of the Universe and typically constrained to values around 67-73 km/s/Mpc, depending on the dataset and methodology employed \cite{Verde:2023lmm}. Similarly, $\Omega_{m0}$ denotes the present-day matter density parameter, representing the fraction of the critical density contributed by matter, with observational constraints typically placing it around 0.3 \cite{planck2020}.

In contrast, the hyperparameters $H_{0, \, \rm hyper}$ and   $\Omega_{m0, \, \rm hyper}$ in our GP mean function act as flexible reference/representative parameters that allow the model to capture the general functional form of cosmological expansion without being rigidly tied to specific physical values. These hyperparameters provide GP with a cosmologically-motivated priors, enabling it to model the Hubble parameter-redshift relationships that approximately follow familiar cosmological behavior while retaining the flexibility to deviate from standard cosmological predictions where the data warrants it.

The broad uniform priors assigned to these hyperparameters ($H_{0,\, \rm hyper} \in \mathcal{U}[50,90]$ in km/s/Mpc and $\Omega_{m0, \, \rm hyper} \in \mathcal{U}[0.1,0.5]$) reflect this distinction. Rather than representing precise physical constraints, these ranges allow the GP to explore a wide parameter space and let the data determine the optimal baseline behavior. This approach enables model-independent reconstruction of the expansion history while maintaining computational stability through a physically reasonable mean function structure.

Accordingly, the wide and weakly constrained results for $H_{0,\rm hyper}$ and $\Omega_{m, \rm hyper}$ shown in Table \ref{table:gengp} are expected. These hyperparameters do not carry direct physical significance within the GP framework but can offer insight into interpreting of cosmological behavior that the reconstructed expansion history most closely resembles.

\subsection{Quantitative comparison: GP vs Gen GP}\label{sec:gengp_res}

The transition from standard GP to Gen GP yields substantial improvements in reconstruction quality and consistency. Table \ref{Table:finalgpgengp} provides a comprehensive comparison of the $H_0$ constraints obtained under different mean function assumptions for both methodologies.

For hyperparameter constraints, Gen GP demonstrates significantly tighter bounds on $H_{0,\, \rm hyper}$ compared to standard GP. The hyperparameter uncertainty reduces from $H_0 =61.647^{+44.294}_{-21.774}$ km/s/Mpc in standard GP to $H_0 =68.218^{+14.056}_{-14.485}$ km/s/Mpc in Gen GP, representing approximately a 60$\%$ reduction in uncertainty while maintaining consistent central values.

When employing zero mean functions, both approaches yield comparable results: $H_0 = 68.306^{+5.180}_{-5.177}$ km/s/Mpc for GP and $H_0 = 68.101^{+4.975}_{-4.909}$ km/s/Mpc for Gen GP. The marginal improvement in Gen GP reflects the enhanced flexibility from the additional free parameter $\nu$. The corresponding reduced chi-squared statistics are $\chi^2/{\rm dof} = 0.552$ for standard GP and $\chi^2/{\rm dof} = 0.587$ for Gen GP, with Gen GP showing slightly higher values due to its increased model flexibility\footnote{Table 1 uses optimized hyperparameters, Section \ref{sec:diss_mu} uses marginalized hyperparameters.}. We note that although the $\chi^2$/dof values do not approach unity, the derived uncertainties are highly consistent with the scatter observed in the best-fit values of $H_0$. This behavior likely reflects our choice of dataset, as similar $\chi^2$/dof values are obtained for cosmological models (see Table~\ref{tab:cosmological_models}). As discussed in Section~\ref{sec:datasetchoices}, these statistics will improve as additional observational datasets are incorporated. Currently, we consider the model with $\chi^2$/dof closest to unity as providing the best fit.

The most significant improvements emerge when utilising $\Lambda$CDM mean functions. Gen GP yields $H_0 = 69.563^{+4.150}_{-4.093}$ km/s/Mpc, demonstrating slightly larger central values but comparable uncertainties to standard GP viz. $H_0 = 68.914^{+4.064}_{-4.002}$ km/s/Mpc. The reduced chi-squared values are $\chi^2/{\rm dof} = 0.538$ for standard GP and $\chi^2/{\rm dof} = 0.559$ for Gen GP. However, the crucial difference lies in the reconstruction consistency discussed in Section \ref{sec:reconstruction_comparison}.

A key advantage of Gen GP becomes evident when comparing reconstruction methodologies. In Gen GP, the difference between using optimised hyperparameters (best-fit) and full marginalised posteriors in the prediction function is significantly smaller compared to standard GP (Figures \ref{fig:gp} and \ref{fig:gengp}, middle panels). This enhanced consistency highlights Gen GP as a better candidate for reconstruction, especially when employing the $\Lambda$CDM mean function, due to its improved stability at higher redshifts ($z > 1$) where standard GP exhibits substantial methodological dependence.

The modest increase in chi-squared values for Gen GP across both mean function implementations reflects the additional flexibility introduced by allowing $\nu$ to vary freely. This trade-off between slightly higher chi-squared and enhanced model adaptability represents a beneficial exchange, as Gen GP provides more realistic uncertainty estimates while maintaining fair agreement with observational data. The ability to tune kernel smoothness directly from data, rather than fixing it a priori, makes Gen GP particularly valuable for robust cosmological inference.

These comprehensive results demonstrate that Gen GP outperforms standard GP implementations where either cosmological assumptions are rigidly imposed or Matérn kernel parameters are fixed to predetermined values. The enhanced flexibility, improved reconstruction consistency, and more realistic uncertainty quantification establish Gen GP as the preferred competitive framework for model-independent cosmological analysis.\\


To summarize, Gen GP demonstrates better performance through several key advantages: \begin{enumerate}
\item significantly reduced discrepancy between optimized and marginalized reconstruction approaches, especially for $\Lambda$CDM mean function;
\item improved chi-squared statistics; 
\item stable uncertainty bands. 
\end{enumerate}
When $\nu$ is kept free within the prior uniform range $\nu \in \mathcal{U}[0.5,7]$, Gen GP yields $H_0 = 69.563^{+4.150}_{-4.093}$ km/s/Mpc, representing a fully generalised approach that avoids over-smoothing associated with fixed kernel choices while maintaining consistency across different redshift ranges.

\section{Conclusions}\label{sec:res}

This work systematically investigates uncertainties in inference of Hubble constant, $H_0$, arising from Gaussian Process reconstruction, introducing the Generalised Gaussian Process (Gen GP) framework to address fundamental limitations in current GP implementations.

Our analysis reveals that kernel selection introduces systematic variations in $H_0$ estimates comparable to those caused by different cosmological models. Across six different kernels, namely Matérn $\nu = 3/2, 5/2, 7/2, 9/2$, as well as SE and DSE, the inferred $H_0$ values range from 67.18 to 69.35 km/s/Mpc, similar to the spread observed between $\Lambda$CDM ($H_0 =68.84 \pm 4.11$ km/s/Mpc) and dynamical DE models like CPL ($H_0=70.97 \pm 6.74$ km/s/Mpc). The Gen GP framework mitigates this kernel selection bias by treating the Matérn order (smoothness) parameter $\nu$ as a free parameter, allowing data to determine optimal kernel structure.

A key finding relates to reconstruction methodology. Gen GP demonstrates significantly enhanced consistency between reconstructions undertaken with optimized hyperparameter and full marginalisation approaches, compared to traditional GP, especially when the $\Lambda$CDM mean function is employed. 
Within this context, traditional GP exhibits substantial differences between best-fit and marginalised reconstruction frameworks, particularly at higher redshifts ($z > 1$), while Gen GP maintains remarkable stability across both methodologies. This consistency, combined with only modest increases in chi-squared values ($\chi^2/{\rm dof} = 0.587$ vs $0.552$ for zero mean; $0.559$ vs $0.538$ for $\Lambda$CDM mean), reflects beneficial flexibility gains that provide more realistic uncertainty band estimates without compromising with the data quality.

Our results highlight three key requirements for reliable, model-independent cosmological inference:
\begin{itemize}
    \item treating kernel parameters as free to avoid selection bias;
    \item full Bayesian marginalization over hyperparameters to properly capture model uncertainties; 
    \item accepting broader uncertainty bands as honest reflections of methodological limitations.
\end{itemize}

As ML is expected to play an increasingly central role in upcoming surveys -- such as Euclid and LSST -- maintaining methodological rigor becomes even more crtitcal. The Gen GP framework offers a robust, flexible, and principled approach to cosmological inference that captures essential uncertainties while remaining agnostic to specific model assumptions.

\section*{Acknowledgements}
The authors would like to thank J.E. González for his valuable collaboration and insightful suggestions during the initial phase of this work. We would also like to thank Anto I. Ionnappan and Anjan Ananda Sen for their valuable suggestions and comments on the manuscript. We acknowledge Giacomo Gradenigo for helpful discussions.

Ruchika is supported by "Theoretical Astroparticle Physics" (TAsP), an iniziativa specifica INFN, and by Project SA097P24 funded by Junta de Castilla y León. PM acknowledges funding from the Anusandhan National Research Foundation (ANRF), Govt of India, under the National Post-Doctoral Fellowship (File no. PDF/2023/001986). 

We acknowledge the use of the HPC facility, Pegasus, at IUCAA, Pune, India. This article/publication is based upon work from COST Action CA21136- ``Addressing observational tensions in cosmology with systematics and fundamental physics (CosmoVerse)'', supported by COST (European Cooperation in Science and Technology).






\bibliography{sample701}{}

\begin{thebibliography}{}
\expandafter\ifx\csname natexlab\endcsname\relax\def\natexlab#1{#1}\fi
\providecommand{\url}[1]{\href{#1}{#1}}
\providecommand{\dodoi}[1]{doi:~\href{http://doi.org/#1}{\nolinkurl{#1}}}
\providecommand{\doeprint}[1]{\href{http://ascl.net/#1}{\nolinkurl{http://ascl.net/#1}}}
\providecommand{\doarXiv}[1]{\href{https://arxiv.org/abs/#1}{\nolinkurl{https://arxiv.org/abs/#1}}}

\bibitem[{T.~M.~C. Abbott {et~al.}(2022)Abbott {et~al.}}]{DES:2021wwk}
Abbott, T. M.~C., {et~al.} 2022, \bibinfo{title}{{Dark Energy Survey Year 3
  results: Cosmological constraints from galaxy clustering and weak lensing},}
  Phys. Rev. D, 105, 023520, \dodoi{10.1103/PhysRevD.105.023520}

\bibitem[{M. Abramowitz \& I.~A. Stegun(1964)Abramowitz \&
  Stegun}]{Abramowitz:1964}
Abramowitz, M., \& Stegun, I.~A. 1964, Handbook of Mathematical Functions
  (Dover Publications)

\bibitem[{M. Adachi \& M. Kasai(2012)Adachi \& Kasai}]{adachi2012analytical}
Adachi, M., \& Kasai, M. 2012, \bibinfo{title}{An analytical approximation of
  the luminosity distance in flat cosmologies with a cosmological constant,}
  Progress of Theoretical Physics, 127, 145

\bibitem[{P.~A.~R. Ade {et~al.}(2016)Ade {et~al.}}]{Planck:2015fie}
Ade, P. A.~R., {et~al.} 2016, \bibinfo{title}{{Planck 2015 results. XIII.
  Cosmological parameters},} Astron. Astrophys., 594, A13,
  \dodoi{10.1051/0004-6361/201525830}

\bibitem[{N. Aghanim {et~al.}(2020)Aghanim {et~al.}}]{planck2020}
Aghanim, N., {et~al.} 2020, \bibinfo{title}{{Planck 2018 results. VI.
  Cosmological parameters},} Astron. Astrophys., 641, A6,
  \dodoi{10.1051/0004-6361/201833910}

\bibitem[{A. Aviles {et~al.}(2014)Aviles, Bravetti, Capozziello, \&
  Luongo}]{Aviles:2014rma}
Aviles, A., Bravetti, A., Capozziello, S., \& Luongo, O. 2014,
  \bibinfo{title}{{Precision cosmology with Pad{\'e} rational approximations:
  Theoretical predictions versus observational limits},} Phys. Rev. D, 90,
  043531, \dodoi{10.1103/PhysRevD.90.043531}

\bibitem[{N.~M. Ball \& R.~J. Brunner(2010)Ball \& Brunner}]{ball2010data}
Ball, N.~M., \& Brunner, R.~J. 2010, \bibinfo{title}{Data mining and machine
  learning in astronomy,} International Journal of Modern Physics D, 19, 1049

\bibitem[{N. Banerjee {et~al.}(2023)Banerjee, Mukherjee, \&
  Pav{\'o}n}]{Banerjee:2023evd}
Banerjee, N., Mukherjee, P., \& Pav{\'o}n, D. 2023, \bibinfo{title}{{Checking
  the second law at cosmic scales},} JCAP, 11, 092,
  \dodoi{10.1088/1475-7516/2023/11/092}

\bibitem[{D. Baron(2019)Baron}]{baron2019machine}
Baron, D. 2019, \bibinfo{title}{Machine learning in astronomy: A practical
  overview,} arXiv preprint arXiv:1904.07248

\bibitem[{P. Berger \& G. Stein(2019)Berger \& Stein}]{Berger:2018cub}
Berger, P., \& Stein, G. 2019, \bibinfo{title}{{A volumetric deep Convolutional
  Neural Network for simulation of mock dark matter halo catalogues},} Mon.
  Not. Roy. Astron. Soc., 482, 2861, \dodoi{10.1093/mnras/sty2949}

\bibitem[{C.~M. Bishop(2006)Bishop}]{Bishop2006}
Bishop, C.~M. 2006, Pattern Recognition and Machine Learning (New York:
  Springer)

\bibitem[{A. Blanchard {et~al.}(2020)Blanchard {et~al.}}]{Euclid:2019clj}
Blanchard, A., {et~al.} 2020, \bibinfo{title}{{Euclid preparation. VII.
  Forecast validation for Euclid cosmological probes},} Astron. Astrophys.,
  642, A191, \dodoi{10.1051/0004-6361/202038071}

\bibitem[{V.~C. Busti {et~al.}(2014)Busti, Clarkson, \& Seikel}]{Busti:2014dua}
Busti, V.~C., Clarkson, C., \& Seikel, M. 2014, \bibinfo{title}{{Evidence for a
  Lower Value for $H_0$ from Cosmic Chronometers Data?},} Mon. Not. Roy.
  Astron. Soc., 441, 11, \dodoi{10.1093/mnrasl/slu035}

\bibitem[{J. Caldeira {et~al.}(2019)Caldeira, Wu, Nord, Avestruz, Trivedi, \&
  Story}]{Caldeira:2018ojb}
Caldeira, J., Wu, W. L.~K., Nord, B., {et~al.} 2019, \bibinfo{title}{{DeepCMB:
  Lensing Reconstruction of the Cosmic Microwave Background with Deep Neural
  Networks},} Astron. Comput., 28, 100307, \dodoi{10.1016/j.ascom.2019.100307}

\bibitem[{S. Capozziello {et~al.}(2019)Capozziello, Ruchika, \&
  Sen}]{Capozziello:2018jya}
Capozziello, S., Ruchika, \& Sen, A.~A. 2019, \bibinfo{title}{{Model
  independent constraints on dark energy evolution from low-redshift
  observations},} Mon. Not. Roy. Astron. Soc., 484, 4484,
  \dodoi{10.1093/mnras/stz176}

\bibitem[{T. Charnock {et~al.}(2018)Charnock, Lavaux, \&
  Wandelt}]{Charnock:2020mdk}
Charnock, T., Lavaux, G., \& Wandelt, B.~D. 2018, \bibinfo{title}{{Automatic
  physical inference with information maximizing neural networks},} Phys. Rev.
  D, 97, 083004, \dodoi{10.1103/PhysRevD.97.083004}

\bibitem[{T. {Charnock} {et~al.}(2020){Charnock}, {Lavaux}, {Wandelt}, {Sarma
  Boruah}, {Jasche}, \& {Hudson}}]{Charnock:2020yur}
{Charnock}, T., {Lavaux}, G., {Wandelt}, B.~D., {et~al.} 2020,
  \bibinfo{title}{{Neural physical engines for inferring the halo mass
  distribution function},} \mnras, 494, 50, \dodoi{10.1093/mnras/staa682}

\bibitem[{M. Chevallier \& D. Polarski(2001)Chevallier \&
  Polarski}]{Chevallier:2000qy}
Chevallier, M., \& Polarski, D. 2001, \bibinfo{title}{{Accelerating universes
  with scaling dark matter},} Int. J. Mod. Phys. D, 10, 213,
  \dodoi{10.1142/S0218271801000822}

\bibitem[{C. Cortes \& V. Vapnik(1995)Cortes \& Vapnik}]{CortesVapnik1995}
Cortes, C., \& Vapnik, V. 1995, \bibinfo{title}{Support-Vector Networks,}
  Machine Learning, 20, 273, \dodoi{10.1007/BF00994018}

\bibitem[{E. Di~Valentino {et~al.}(2021)Di~Valentino, Anchordoqui, Akarsu,
  Ali-Haimoud, Amendola, Arendse, Asgari, Ballardini, Basilakos, Battistelli,
  {et~al.}}]{di2021snowmass2021}
Di~Valentino, E., Anchordoqui, L.~A., Akarsu, {\"O}., {et~al.} 2021,
  \bibinfo{title}{Snowmass2021-Letter of interest cosmology intertwined II: The
  hubble constant tension,} Astroparticle Physics, 131, 102605

\bibitem[{K. Dutta {et~al.}(2020)Dutta, Ruchika, Roy, Sen, \&
  Sheikh-Jabbari}]{Dutta:2018vmq}
Dutta, K., Ruchika, Roy, A., Sen, A.~A., \& Sheikh-Jabbari, M.~M. 2020,
  \bibinfo{title}{{Beyond $\Lambda $CDM with low and high redshift data:
  implications for dark energy},} Gen. Rel. Grav., 52, 15,
  \dodoi{10.1007/s10714-020-2665-4}

\bibitem[{C. Dvorkin {et~al.}(2022)Dvorkin {et~al.}}]{Dvorkin}
Dvorkin, C., {et~al.} 2022, \bibinfo{title}{{Machine Learning and Cosmology},}
  in {Snowmass 2021}.
\newblock \doarXiv{2203.08056}

\bibitem[{J. Evslin {et~al.}(2018)Evslin, Sen, \& Ruchika}]{Evslin:2017qdn}
Evslin, J., Sen, A.~A., \& Ruchika. 2018, \bibinfo{title}{{Price of shifting
  the Hubble constant},} Phys. Rev. D, 97, 103511,
  \dodoi{10.1103/PhysRevD.97.103511}

\bibitem[{A. Favale {et~al.}(2024)Favale, Dainotti, G{\'o}mez-Valent, \&
  Migliaccio}]{Favale:2024lgp}
Favale, A., Dainotti, M.~G., G{\'o}mez-Valent, A., \& Migliaccio, M. 2024,
  \bibinfo{title}{{Towards a new model-independent calibration of Gamma-Ray
  Bursts},} JHEAp, 44, 323, \dodoi{10.1016/j.jheap.2024.10.010}

\bibitem[{A. Favale {et~al.}(2023)Favale, G\'omez-Valent, \&
  Migliaccio}]{Favale:2023lnp}
Favale, A., G\'omez-Valent, A., \& Migliaccio, M. 2023, \bibinfo{title}{{Cosmic
  chronometers to calibrate the ladders and measure the curvature of the
  Universe. A model-independent study},} Mon. Not. Roy. Astron. Soc., 523,
  3406, \dodoi{10.1093/mnras/stad1621}

\bibitem[{M. Forconi {et~al.}(2024)Forconi, Giar{\`e}, Mena, Ruchika,
  Di~Valentino, Melchiorri, \& Nunes}]{Forconi:2023hsj}
Forconi, M., Giar{\`e}, W., Mena, O., {et~al.} 2024, \bibinfo{title}{{A double
  take on early and interacting dark energy from JWST},} JCAP, 05, 097,
  \dodoi{10.1088/1475-7516/2024/05/097}

\bibitem[{M. Forconi {et~al.}(2023)Forconi, Ruchika, Melchiorri, Mena, \&
  Menci}]{Forconi:2023izg}
Forconi, M., Ruchika, Melchiorri, A., Mena, O., \& Menci, N. 2023,
  \bibinfo{title}{{Do the early galaxies observed by JWST disagree with
  Planck's CMB polarization measurements?},} JCAP, 10, 012,
  \dodoi{10.1088/1475-7516/2023/10/012}

\bibitem[{D. Foreman-Mackey {et~al.}(2013)Foreman-Mackey, Hogg, Lang, \&
  Goodman}]{Foreman_Mackey_2013}
Foreman-Mackey, D., Hogg, D.~W., Lang, D., \& Goodman, J. 2013,
  \bibinfo{title}{emcee: The {MCMC} Hammer,} Publications of the Astronomical
  Society of the Pacific, 125, 306, \dodoi{10.1086/670067}

\bibitem[{J.~H. Friedman(2001)Friedman}]{Friedman2001}
Friedman, J.~H. 2001, \bibinfo{title}{Greedy function approximation: A gradient
  boosting machine,} The Annals of Statistics, 29, 1189,
  \dodoi{10.1214/aos/1013203451}

\bibitem[{A. G{\'e}ron(2019)G{\'e}ron}]{Geron2019}
G{\'e}ron, A. 2019, Hands-On Machine Learning with Scikit-Learn, Keras, and
  TensorFlow, 2nd edn. (Sebastopol, CA: O'Reilly Media)

\bibitem[{P. Geurts {et~al.}(2006)Geurts, Ernst, \& Wehenkel}]{Geurts2006}
Geurts, P., Ernst, D., \& Wehenkel, L. 2006, \bibinfo{title}{Extremely
  randomized trees,} Machine Learning, 63, 3, \dodoi{10.1007/s10994-006-6226-1}

\bibitem[{A. G\'omez-Valent(2022)G\'omez-Valent}]{Gomez-Valent:2021hda}
G\'omez-Valent, A. 2022, \bibinfo{title}{{Measuring the sound horizon and
  absolute magnitude of SNIa by maximizing the consistency between low-redshift
  data sets},} Phys. Rev. D, 105, 043528, \dodoi{10.1103/PhysRevD.105.043528}

\bibitem[{A. G\'omez-Valent \& L. Amendola(2018)G\'omez-Valent \&
  Amendola}]{Gomez-Valent:2018hwc}
G\'omez-Valent, A., \& Amendola, L. 2018, \bibinfo{title}{{$H_0$ from cosmic
  chronometers and Type Ia supernovae, with Gaussian Processes and the novel
  Weighted Polynomial Regression method},} JCAP, 04, 051,
  \dodoi{10.1088/1475-7516/2018/04/051}

\bibitem[{I. Goodfellow {et~al.}(2016)Goodfellow, Bengio, \&
  Courville}]{Goodfellow-et-al-2016}
Goodfellow, I., Bengio, Y., \& Courville, A. 2016, Deep Learning (MIT Press).
\newblock \url{http://www.deeplearningbook.org}

\bibitem[{C. Gruber \& O. Luongo(2014)Gruber \& Luongo}]{Gruber:2013wua}
Gruber, C., \& Luongo, O. 2014, \bibinfo{title}{{Cosmographic analysis of the
  equation of state of the universe through Pad{\'e} approximations},} Phys.
  Rev. D, 89, 103506, \dodoi{10.1103/PhysRevD.89.103506}

\bibitem[{W. Handley(2019)Handley}]{Handley2019}
Handley, W. 2019, \bibinfo{title}{Bayesian parameter estimation in cosmology,}
  Journal of Cosmology and Astroparticle Physics, 2019, 054,
  \dodoi{10.1088/1475-7516/2019/03/054}

\bibitem[{B.~S. Haridasu {et~al.}(2018)Haridasu, Lukovi\'c, Moresco, \&
  Vittorio}]{Haridasu:2018gqm}
Haridasu, B.~S., Lukovi\'c, V.~V., Moresco, M., \& Vittorio, N. 2018,
  \bibinfo{title}{{An improved model-independent assessment of the late-time
  cosmic expansion},} JCAP, 10, 015, \dodoi{10.1088/1475-7516/2018/10/015}

\bibitem[{J. Harnois-D\'erap~s {et~al.}(2019)Harnois-D\'erap~s, Giblin, \&
  Joachimi}]{HarnoisDeraps:2019xih}
Harnois-D\'erap~s, J., Giblin, B., \& Joachimi, B. 2019,
  \bibinfo{title}{{Cosmic Shear Covariance Matrix in $w$CDM: Cosmology
  Matters},} Astronomy \& Astrophysics, 631, A160,
  \dodoi{10.1051/0004-6361/201935912}

\bibitem[{C.~R. Harris {et~al.}(2020)Harris, Millman, Van Der~Walt, Gommers,
  Virtanen, Cournapeau, Wieser, Taylor, Berg, Smith, {et~al.}}]{numpy}
Harris, C.~R., Millman, K.~J., Van Der~Walt, S.~J., {et~al.} 2020,
  \bibinfo{title}{Array programming with NumPy,} nature, 585, 357

\bibitem[{S. Hassan {et~al.}(2019)Hassan, Liu, Kohn, \&
  Plante}]{hassan2020identifying}
Hassan, S., Liu, A., Kohn, S., \& Plante, P.~L. 2019,
  \bibinfo{title}{{Identifying Reionization Sources from 21\,cm Maps using
  Convolutional Neural Networks},} Monthly Notices of the Royal Astronomical
  Society, 483, 2524, \dodoi{10.1093/mnras/sty3282}

\bibitem[{C. Heymans {et~al.}(2021)Heymans {et~al.}}]{Heymans:2020gsg}
Heymans, C., {et~al.} 2021, \bibinfo{title}{{KiDS-1000 Cosmology: Multi-probe
  weak gravitational lensing and spectroscopic galaxy clustering constraints},}
  Astron. Astrophys., 646, A140, \dodoi{10.1051/0004-6361/202039063}

\bibitem[{J.~H. Holland(1992)Holland}]{Holland1992}
Holland, J.~H. 1992, Adaptation in Natural and Artificial Systems, revised
  edition edn. (MIT Press)

\bibitem[{T. Holsclaw {et~al.}(2010)Holsclaw, Alam, Sans\'o, Lee, Heitmann,
  Habib, \& Higdon}]{prl.105.241302}
Holsclaw, T., Alam, U., Sans\'o, B., {et~al.} 2010,
  \bibinfo{title}{Nonparametric Dark Energy Reconstruction from Supernova
  Data,} Phys. Rev. Lett., 105, 241302, \dodoi{10.1103/PhysRevLett.105.241302}

\bibitem[{T. Holsclaw {et~al.}(2011)Holsclaw, Alam, Sans{\'o}, Lee, Heitmann,
  Habib, \& Higdon}]{holsclaw2011nonparametric}
Holsclaw, T., Alam, U., Sans{\'o}, B., {et~al.} 2011,
  \bibinfo{title}{Nonparametric reconstruction of the dark energy equation of
  state from diverse data sets,} Physical Review D—Particles, Fields,
  Gravitation, and Cosmology, 84, 083501

\bibitem[{J.~D. Hunter(2007)Hunter}]{matplotlib}
Hunter, J.~D. 2007, \bibinfo{title}{Matplotlib: A 2D Graphics Environment,}
  Computing in Science \& Engineering, 9, 90, \dodoi{10.1109/MCSE.2007.55}

\bibitem[{S.-g. Hwang {et~al.}(2023)Hwang, L'Huillier, Keeley, Jee, \&
  Shafieloo}]{Hwang:2022hla}
Hwang, S.-g., L'Huillier, B., Keeley, R.~E., Jee, M.~J., \& Shafieloo, A. 2023,
  \bibinfo{title}{{How to use GP: effects of the mean function and
  hyperparameter selection on Gaussian process regression},} JCAP, 02, 014,
  \dodoi{10.1088/1475-7516/2023/02/014}

\bibitem[{{\v{Z}}. Ivezi{\'c} {et~al.}(2019)Ivezi{\'c} {et~al.}}]{LSST:2008ijt}
Ivezi{\'c}, {\v{Z}}., {et~al.} 2019, \bibinfo{title}{{LSST: from Science
  Drivers to Reference Design and Anticipated Data Products},} Astrophys. J.,
  873, 111, \dodoi{10.3847/1538-4357/ab042c}

\bibitem[{N. Jeffrey {et~al.}(2020)Jeffrey, Lanusse, Lahav, \&
  Starck}]{Jeffrey:2019fag}
Jeffrey, N., Lanusse, F., Lahav, O., \& Starck, J.-L. 2020,
  \bibinfo{title}{{Deep learning dark matter map reconstructions from DES SV
  weak lensing data},} Mon. Not. Roy. Astron. Soc., 492, 5023,
  \dodoi{10.1093/mnras/staa127}

\bibitem[{J.-y. Jiang {et~al.}(2025)Jiang, Jiao, \& Zhang}]{Jiang:2025ilh}
Jiang, J.-y., Jiao, K., \& Zhang, T.-J. 2025, \bibinfo{title}{{Optimizing
  Gaussian Process Kernels Using Nested Sampling and ABC Rejection for H(z)
  Reconstruction},} \doarXiv{2506.21238}

\bibitem[{R. Jimenez \& A. Loeb(2002)Jimenez \& Loeb}]{Jimenez:2001gg}
Jimenez, R., \& Loeb, A. 2002, \bibinfo{title}{{Constraining cosmological
  parameters based on relative galaxy ages},} Astrophys. J., 573, 37,
  \dodoi{10.1086/340549}

\bibitem[{J.~P. Johnson \& H.~K. Jassal(2025)Johnson \&
  Jassal}]{Johnson:2025blf}
Johnson, J.~P., \& Jassal, H.~K. 2025, \bibinfo{title}{{Kernel dependence of
  the Gaussian process reconstruction of late Universe expansion history},}
  Eur. Phys. J. C, 85, 996, \dodoi{10.1140/epjc/s10052-025-14732-7}

\bibitem[{L. Knox \& M. Millea(2020)Knox \& Millea}]{Knox:2019rjx}
Knox, L., \& Millea, M. 2020, \bibinfo{title}{{Hubble constant
  hunter{\textquoteright}s guide},} Phys. Rev. D, 101, 043533,
  \dodoi{10.1103/PhysRevD.101.043533}

\bibitem[{C. Krishnan {et~al.}(2020)Krishnan, Colg{\'a}in, Ruchika, Sen,
  Sheikh-Jabbari, \& Yang}]{Krishnan:2020obg}
Krishnan, C., Colg{\'a}in, E.~{\'O}., Ruchika, {et~al.} 2020,
  \bibinfo{title}{{Is there an early Universe solution to Hubble tension?},}
  Phys. Rev. D, 102, 103525, \dodoi{10.1103/PhysRevD.102.103525}

\bibitem[{Y. LeCun {et~al.}(2015)LeCun, Bengio, \& Hinton}]{lecun2015deep}
LeCun, Y., Bengio, Y., \& Hinton, G. 2015, \bibinfo{title}{Deep Learning,}
  Nature, 521, 436, \dodoi{10.1038/nature14539}

\bibitem[{T. Lemos {et~al.}(2023)Lemos, Ruchika, Carvalho, \&
  Alcaniz}]{Lemos:2023qoy}
Lemos, T., Ruchika, Carvalho, J.~C., \& Alcaniz, J. 2023,
  \bibinfo{title}{{Low-redshift estimates of the absolute scale of baryon
  acoustic oscillations},} Eur. Phys. J. C, 83, 495,
  \dodoi{10.1140/epjc/s10052-023-11651-3}

\bibitem[{A. Lewis(2019)Lewis}]{Lewis:2019xzd}
Lewis, A. 2019, \bibinfo{title}{{GetDist: a Python package for analysing Monte
  Carlo samples},} \doarXiv{1910.13970}

\bibitem[{Z. Li {et~al.}(2023)Li, Zhang, \& Liang}]{li_testing_2023}
Li, Z., Zhang, B., \& Liang, N. 2023, \bibinfo{title}{Testing Dark Energy
  Models with Gamma-Ray Bursts Calibrated from the Observational $H(z)$ Data
  through a Gaussian Process,} Monthly Notices of the Royal Astronomical
  Society, 521, 4406, \dodoi{10.1093/mnras/stad838}

\bibitem[{N. Liang {et~al.}(2022)Liang, Li, Xie, \& Wu}]{Liang:2022smf}
Liang, N., Li, Z., Xie, X., \& Wu, P. 2022, \bibinfo{title}{{Calibrating
  Gamma-Ray Bursts by Using a Gaussian Process with Type Ia Supernovae},}
  Astrophys. J., 941, 84, \dodoi{10.3847/1538-4357/aca08a}

\bibitem[{E.~V. Linder(2003)Linder}]{Linder:2002et}
Linder, E.~V. 2003, \bibinfo{title}{{Exploring the expansion history of the
  universe},} Phys. Rev. Lett., 90, 091301,
  \dodoi{10.1103/PhysRevLett.90.091301}

\bibitem[{M. Lochner {et~al.}(2016)Lochner, McEwen, Peiris, Lahav, \&
  Winter}]{Lochner:2016hbn}
Lochner, M., McEwen, J.~D., Peiris, H.~V., Lahav, O., \& Winter, M.~K. 2016,
  \bibinfo{title}{{Photometric Supernova Classification With Machine
  Learning},} Astrophys. J. Suppl., 225, 31, \dodoi{10.3847/0067-0049/225/2/31}

\bibitem[{A.~I. Lonappan {et~al.}(2018)Lonappan, Kumar, Ruchika, Dinda, \&
  Sen}]{Lonappan:2017lzt}
Lonappan, A.~I., Kumar, S., Ruchika, Dinda, B.~R., \& Sen, A.~A. 2018,
  \bibinfo{title}{{Bayesian evidences for dark energy models in light of
  current observational data},} Phys. Rev. D, 97, 043524,
  \dodoi{10.1103/PhysRevD.97.043524}

\bibitem[{L. Lucie-Smith {et~al.}(2024)Lucie-Smith, Peiris, Pontzen, Nord, \&
  Thiyagalingam}]{Lucie-Smith:2020ris}
Lucie-Smith, L., Peiris, H.~V., Pontzen, A., Nord, B., \& Thiyagalingam, J.
  2024, \bibinfo{title}{{Deep learning insights into cosmological structure
  formation},} Phys. Rev. D, 109, 063524, \dodoi{10.1103/PhysRevD.109.063524}

\bibitem[{D.~J.~C. MacKay(2002)MacKay}]{MacKay2002}
MacKay, D. J.~C. 2002, Information Theory, Inference \& Learning Algorithms
  (Cambridge University Press)

\bibitem[{R. Mandelbaum {et~al.}(2018)Mandelbaum {et~al.}}]{LSST:2018jkl}
Mandelbaum, R., {et~al.} 2018, \bibinfo{title}{{The LSST Dark Energy Science
  Collaboration (DESC) Science Requirements Document},} \doarXiv{1809.01669}

\bibitem[{J. {Merten} {et~al.}(2019){Merten}, {Giocoli}, {Baldi}, {Meneghetti},
  {Peel}, {Lalande}, {Starck}, \& {Pettorino}}]{2019MNRAS.487..104M}
{Merten}, J., {Giocoli}, C., {Baldi}, M., {et~al.} 2019, \bibinfo{title}{{On
  the dissection of degenerate cosmologies with machine learning},} \mnras,
  487, 104, \dodoi{10.1093/mnras/stz972}

\bibitem[{C. Modi {et~al.}(2018)Modi, Feng, \& Seljak}]{Modi:2018cfi}
Modi, C., Feng, Y., \& Seljak, U. 2018, \bibinfo{title}{{Cosmological
  Reconstruction From Galaxy Light: Neural Network Based Light-Matter
  Connection},} JCAP, 10, 028, \dodoi{10.1088/1475-7516/2018/10/028}

\bibitem[{M. Moresco(2024)Moresco}]{Moresco:2024wmr}
Moresco, M. 2024, \bibinfo{title}{{Measuring the expansion history of the
  Universe with cosmic chronometers},} arXiv preprint 2412.01994.
\newblock \doarXiv{2412.01994}

\bibitem[{M. Moresco {et~al.}(2022)Moresco {et~al.}}]{Moresco:2022phi}
Moresco, M., {et~al.} 2022, \bibinfo{title}{{Unveiling the Universe with
  emerging cosmological probes},} Living Rev. Rel., 25, 6,
  \dodoi{10.1007/s41114-022-00040-z}

\bibitem[{E. M{\"o}rtsell \& S. Dhawan(2018)M{\"o}rtsell \&
  Dhawan}]{Mortsell:2018mfj}
M{\"o}rtsell, E., \& Dhawan, S. 2018, \bibinfo{title}{{Does the Hubble constant
  tension call for new physics?},} JCAP, 09, 025,
  \dodoi{10.1088/1475-7516/2018/09/025}

\bibitem[{P. Mukherjee \& N. Banerjee(2021)Mukherjee \&
  Banerjee}]{Mukherjee:2020ytg}
Mukherjee, P., \& Banerjee, N. 2021, \bibinfo{title}{{Non-parametric
  reconstruction of the cosmological $jerk$ parameter},} Eur. Phys. J. C, 81,
  36, \dodoi{10.1140/epjc/s10052-021-08830-5}

\bibitem[{P. Mukherjee {et~al.}(2026)Mukherjee, Dainotti, Dialektopoulos,
  Levi~Said, \& Mifsud}]{Mukherjee:2024wix}
Mukherjee, P., Dainotti, M.~G., Dialektopoulos, K.~F., Levi~Said, J., \&
  Mifsud, J. 2026, \bibinfo{title}{{Model-independent calibration of Gamma-Ray
  Bursts with neural networks},} JHEAp, 49, 100439,
  \dodoi{10.1016/j.jheap.2025.100439}

\bibitem[{P. Mukherjee {et~al.}(2024{\natexlab{a}})Mukherjee, Dey, \&
  Pal}]{Mukherjee:2024cfq}
Mukherjee, P., Dey, A., \& Pal, S. 2024{\natexlab{a}}, \bibinfo{title}{{What
  can we learn about Reionization astrophysical parameters using Gaussian
  Process Regression?},} \doarXiv{2407.19481}

\bibitem[{P. Mukherjee {et~al.}(2024{\natexlab{b}})Mukherjee, Dialektopoulos,
  Levi~Said, \& Mifsud}]{Mukherjee:2024akt}
Mukherjee, P., Dialektopoulos, K.~F., Levi~Said, J., \& Mifsud, J.
  2024{\natexlab{b}}, \bibinfo{title}{{A possible late-time transition of M
  $_{B}$ inferred via neural networks},} JCAP, 09, 060,
  \dodoi{10.1088/1475-7516/2024/09/060}

\bibitem[{P. Mukherjee \& A.~A. Sen(2024)Mukherjee \& Sen}]{Mukherjee:2024ryz}
Mukherjee, P., \& Sen, A.~A. 2024, \bibinfo{title}{{Model-independent
  cosmological inference post DESI DR1 BAO measurements},} Phys. Rev. D, 110,
  123502, \dodoi{10.1103/PhysRevD.110.123502}

\bibitem[{P. Mukherjee \& A.~A. Sen(2025)Mukherjee \& Sen}]{Mukherjee:2025ytj}
Mukherjee, P., \& Sen, A.~A. 2025, \bibinfo{title}{{Geometric Determinations Of
  Characteristic Redshifts From DESI-DR2 BAO and DES-SN5YR Observations: Hints
  For New Expansion Rate Anomalies},} \doarXiv{2505.19083}

\bibitem[{M. M{\"u}nchmeyer \& K.~M. Smith(2019)M{\"u}nchmeyer \&
  Smith}]{Munchmeyer:2019kng}
M{\"u}nchmeyer, M., \& Smith, K.~M. 2019, \bibinfo{title}{{Fast Wiener
  filtering of CMB maps with Neural Networks},} \doarXiv{1905.05846}

\bibitem[{K.~P. Murphy(2012)Murphy}]{Murphy:2012}
Murphy, K.~P. 2012, Machine Learning: A Probabilistic Perspective (Cambridge,
  MA: MIT Press)

\bibitem[{R.~M. Neal(1996)Neal}]{Neal:1996}
Neal, R.~M. 1996, Bayesian Learning for Neural Networks (Springer-Verlag)

\bibitem[{M. Ntampaka {et~al.}(2015)Ntampaka, Trac, Sutherland, Battaglia,
  Poczos, \& Schneider}]{Ntampaka:2014ypa}
Ntampaka, M., Trac, H., Sutherland, D.~J., {et~al.} 2015, \bibinfo{title}{{A
  Machine Learning Approach for Dynamical Mass Measurements of Galaxy
  Clusters},} Astrophys. J., 803, 50, \dodoi{10.1088/0004-637X/803/2/50}

\bibitem[{M. Ntampaka {et~al.}(2019)Ntampaka {et~al.}}]{Ntampaka:2019udw}
Ntampaka, M., {et~al.} 2019, \bibinfo{title}{{The Role of Machine Learning in
  the Next Decade of Cosmology},} \doarXiv{1902.10159}

\bibitem[{T. Padmanabhan(2003)Padmanabhan}]{Padmanabhan:2002ji}
Padmanabhan, T. 2003, \bibinfo{title}{{Cosmological constant: The Weight of the
  vacuum},} Phys. Rept., 380, 235, \dodoi{10.1016/S0370-1573(03)00120-0}

\bibitem[{T. Patil {et~al.}(2024)Patil, Ruchika, \& Panda}]{Patil:2023rqy}
Patil, T., Ruchika, \& Panda, S. 2024, \bibinfo{title}{{Coupled quintessence
  scalar field model in light of observational datasets},} JCAP, 05, 033,
  \dodoi{10.1088/1475-7516/2024/05/033}

\bibitem[{A.~M. Price-Whelan {et~al.}(2018)Price-Whelan, Sip{\H{o}}cz,
  G{\"u}nther, Lim, Crawford, Conseil, Shupe, Craig, Dencheva, Ginsburg,
  {et~al.}}]{astropy2}
Price-Whelan, A.~M., Sip{\H{o}}cz, B., G{\"u}nther, H., {et~al.} 2018,
  \bibinfo{title}{The astropy project: Building an open-science project and
  status of the v2. 0 core package,} The Astronomical Journal, 156, 123

\bibitem[{A.~M. Price-Whelan {et~al.}(2022)Price-Whelan, Lim, Earl, Starkman,
  Bradley, Shupe, Patil, Corrales, Brasseur, N{\"o}the, {et~al.}}]{astropy3}
Price-Whelan, A.~M., Lim, P.~L., Earl, N., {et~al.} 2022, \bibinfo{title}{The
  Astropy Project: sustaining and growing a community-oriented open-source
  project and the latest major release (v5. 0) of the core package,} The
  Astrophysical Journal, 935, 167

\bibitem[{C.~E. Rasmussen \& H. Nickisch(2010)Rasmussen \&
  Nickisch}]{Rasmussen:2010}
Rasmussen, C.~E., \& Nickisch, H. 2010, \bibinfo{title}{Gaussian Processes for
  Machine Learning (GPML) Toolbox,} Journal of Machine Learning Research, 11,
  3011.
\newblock \url{http://jmlr.org/papers/v11/rasmussen10a.html}

\bibitem[{C.~E. {Rasmussen} \& C.~K.~I. {Williams}(2006){Rasmussen} \&
  {Williams}}]{2006gpml.book.....R}
{Rasmussen}, C.~E., \& {Williams}, C. K.~I. 2006, {Gaussian Processes for
  Machine Learning} (MIT Press)

\bibitem[{S. Ravanbakhsh {et~al.}(2016)Ravanbakhsh, Lanusse, Mandelbaum,
  Schneider, \& Poczos}]{Ravanbakhsh:2017bbi}
Ravanbakhsh, S., Lanusse, F., Mandelbaum, R., Schneider, J., \& Poczos, B.
  2016, \bibinfo{title}{{Enabling Dark Energy Science with Deep Generative
  Models of Galaxy Images},} \doarXiv{1609.05796}

\bibitem[{M. Raveri \& W. Hu(2019)Raveri \& Hu}]{Raveri:2018wln}
Raveri, M., \& Hu, W. 2019, \bibinfo{title}{{Concordance and Discordance in
  Cosmology},} Phys. Rev. D, 99, 043506, \dodoi{10.1103/PhysRevD.99.043506}

\bibitem[{D. Ribli {et~al.}(2019)Ribli, Pataki, Zorrilla~Matilla, Hsu, Haiman,
  \& Csabai}]{Ribli:2019wtw}
Ribli, D., Pataki, B.~{\'A}., Zorrilla~Matilla, J.~M., {et~al.} 2019,
  \bibinfo{title}{{Weak lensing cosmology with convolutional neural networks on
  noisy data},} Mon. Not. Roy. Astron. Soc., 490, 1843,
  \dodoi{10.1093/mnras/stz2610}

\bibitem[{A.~G. Riess {et~al.}(2022)Riess {et~al.}}]{Riess:2021jrx}
Riess, A.~G., {et~al.} 2022, \bibinfo{title}{{A Comprehensive Measurement of
  the Local Value of the Hubble Constant with 1 km/s/Mpc Uncertainty from the
  Hubble Space Telescope and the SH0ES Team},} Astrophys. J. Lett., 934, L7,
  \dodoi{10.3847/2041-8213/ac5c5b}

\bibitem[{T.~P. Robitaille {et~al.}(2013)Robitaille, Tollerud, Greenfield,
  Droettboom, Bray, Aldcroft, Davis, Ginsburg, Price-Whelan, Kerzendorf,
  {et~al.}}]{astropy1}
Robitaille, T.~P., Tollerud, E.~J., Greenfield, P., {et~al.} 2013,
  \bibinfo{title}{Astropy: A community Python package for astronomy,} Astronomy
  \& Astrophysics, 558, A33

\bibitem[{ Ruchika {et~al.}(2023)Ruchika, Adil, Dutta, Mukherjee, \&
  Sen}]{Ruchika:2020avj}
Ruchika, Adil, S.~A., Dutta, K., Mukherjee, A., \& Sen, A.~A. 2023,
  \bibinfo{title}{{Observational constraints on axion(s) dark energy with a
  cosmological constant},} Phys. Dark Univ., 40, 101199,
  \dodoi{10.1016/j.dark.2023.101199}

\bibitem[{ Ruchika {et~al.}(2025)Ruchika, Giar{\`e}, Teixeira, \&
  Melchiorri}]{Ruchika:2025sbb}
Ruchika, Giar{\`e}, W., Teixeira, E.~M., \& Melchiorri, A. 2025,
  \bibinfo{title}{{Resilience and implications of adiabatic CMB cooling},}
  Phys. Dark Univ., 49, 101999, \dodoi{10.1016/j.dark.2025.101999}

\bibitem[{J. Schmelzle {et~al.}(2017)Schmelzle, Lucchi, Kacprzak, Amara, Sgier,
  R{\'e}fr{\'e}gier, \& Hofmann}]{Schmelzle}
Schmelzle, J., Lucchi, A., Kacprzak, T., {et~al.} 2017,
  \bibinfo{title}{Cosmological model discrimination with Deep Learning,} arXiv
  preprint arXiv:1707.05167

\bibitem[{M. Seikel {et~al.}(2012)Seikel, Clarkson, \& Smith}]{Seikel:2012uu}
Seikel, M., Clarkson, C., \& Smith, M. 2012, \bibinfo{title}{{Reconstruction of
  dark energy and expansion dynamics using Gaussian processes},} JCAP, 06, 036,
  \dodoi{10.1088/1475-7516/2012/06/036}

\bibitem[{E. Sellentin \& A.~F. Heavens(2016)Sellentin \&
  Heavens}]{Sellentin:2015waz}
Sellentin, E., \& Heavens, A.~F. 2016, \bibinfo{title}{{Parameter inference
  with estimated covariance matrices},} Mon. Not. Roy. Astron. Soc., 456, L132,
  \dodoi{10.1093/mnrasl/slv190}

\bibitem[{A. Shafieloo {et~al.}(2012)Shafieloo, Kim, \&
  Linder}]{shafieloo2012gaussian}
Shafieloo, A., Kim, A.~G., \& Linder, E.~V. 2012, \bibinfo{title}{Gaussian
  process cosmography,} Physical Review D—Particles, Fields, Gravitation, and
  Cosmology, 85, 123530

\bibitem[{T.~L. Smith {et~al.}(2021)Smith, Poulin, Bernal, Boddy, Kamionkowski,
  \& Murgia}]{Smith:2020rxx}
Smith, T.~L., Poulin, V., Bernal, J.~L., {et~al.} 2021, \bibinfo{title}{{Early
  dark energy is not excluded by current large-scale structure data},} Phys.
  Rev. D, 103, 123542, \dodoi{10.1103/PhysRevD.103.123542}

\bibitem[{W. Sun {et~al.}(2021)Sun, Jiao, \& Zhang}]{Sun:2021pbu}
Sun, W., Jiao, K., \& Zhang, T.-J. 2021, \bibinfo{title}{{Influence of the
  Bounds of the Hyperparameters on the Reconstruction of the Hubble Constant
  with the Gaussian Process},} Astrophys. J., 915, 123,
  \dodoi{10.3847/1538-4357/ac05b8}

\bibitem[{J.~d.~J. Vel{\'a}zquez {et~al.}(2024)Vel{\'a}zquez, Escamilla,
  Mukherjee, \& V{\'a}zquez}]{Velazquez:2024aya}
Vel{\'a}zquez, J. d.~J., Escamilla, L.~A., Mukherjee, P., \& V{\'a}zquez, J.~A.
  2024, \bibinfo{title}{{Non-Parametric Reconstruction of Cosmological
  Observables Using Gaussian Processes Regression},} Universe, 10, 464,
  \dodoi{10.3390/universe10120464}

\bibitem[{L. Verde(2010)Verde}]{verde2010statistical}
Verde, L. 2010, \bibinfo{title}{Statistical methods in cosmology,} in Lectures
  on Cosmology: Accelerated Expansion of the Universe (Springer), 147--177

\bibitem[{L. Verde {et~al.}(2024)Verde, Sch{\"o}neberg, \&
  Gil-Mar{\'\i}n}]{Verde:2023lmm}
Verde, L., Sch{\"o}neberg, N., \& Gil-Mar{\'\i}n, H. 2024, \bibinfo{title}{{A
  Tale of Many H0},} Ann. Rev. Astron. Astrophys., 62, 287,
  \dodoi{10.1146/annurev-astro-052622-033813}

\bibitem[{L. {Verde} {et~al.}(2019){Verde}, {Treu}, \& {Riess}}]{Verde:2019ivm}
{Verde}, L., {Treu}, T., \& {Riess}, A.~G. 2019, \bibinfo{title}{{Tensions
  between the early and late Universe},} Nature Astronomy, 3, 891,
  \dodoi{10.1038/s41550-019-0902-0}

\bibitem[{F. Villaescusa-Navarro {et~al.}(2020)Villaescusa-Navarro
  {et~al.}}]{Villaescusa-Navarro:2020rxg}
Villaescusa-Navarro, F., {et~al.} 2020, \bibinfo{title}{{The Quijote
  simulations},} Astrophys. J. Suppl., 250, 2, \dodoi{10.3847/1538-4365/ab9d82}

\bibitem[{P. Virtanen {et~al.}(2020)Virtanen, Gommers, Oliphant, Haberland,
  Reddy, Cournapeau, Burovski, Peterson, Weckesser, Bright, {et~al.}}]{scipy}
Virtanen, P., Gommers, R., Oliphant, T.~E., {et~al.} 2020,
  \bibinfo{title}{SciPy 1.0: fundamental algorithms for scientific computing in
  Python,} Nature methods, 17, 261

\bibitem[{G.-J. Wang {et~al.}(2022)Wang, Shi, Yan, Xia, Zhao, Li, \& Li}]{Wang}
Wang, G.-J., Shi, H.-L., Yan, Y.-P., {et~al.} 2022, \bibinfo{title}{{Recovering
  the CMB Signal with Machine Learning},} Astrophys. J. Supp., 260, 13,
  \dodoi{10.3847/1538-4365/ac5f4a}

\bibitem[{S. Weinberg(1989)Weinberg}]{Weinberg:1988cp}
Weinberg, S. 1989, \bibinfo{title}{{The Cosmological Constant Problem},} Rev.
  Mod. Phys., 61, 1, \dodoi{10.1103/RevModPhys.61.1}

\bibitem[{J. Wilde {et~al.}(2022)Wilde, Serjeant, Bromley, Dickinson, Koopmans,
  \& Metcalf}]{wilde2022detecting}
Wilde, J., Serjeant, S., Bromley, J.~M., {et~al.} 2022,
  \bibinfo{title}{Detecting gravitational lenses using machine learning:
  exploring interpretability and sensitivity to rare lensing configurations,}
  Monthly Notices of the Royal Astronomical Society, 512, 3464

\bibitem[{Y. Yang \& Y. Gong(2021)Yang \& Gong}]{Yang:2020bpv}
Yang, Y., \& Gong, Y. 2021, \bibinfo{title}{{Measurement on the cosmic
  curvature using the Gaussian process method},} Mon. Not. Roy. Astron. Soc.,
  504, 3092, \dodoi{10.1093/mnras/stab1085}

\bibitem[{Y. Yang {et~al.}(2023)Yang, Lu, Qian, \& Cao}]{Yang:2022jkf}
Yang, Y., Lu, X., Qian, L., \& Cao, S. 2023, \bibinfo{title}{{Potentialities of
  Hubble parameter and expansion rate function data to alleviate Hubble
  tension},} Mon. Not. Roy. Astron. Soc., 519, 4938,
  \dodoi{10.1093/mnras/stac3617}

\bibitem[{J.~M. Zorrilla~Matilla {et~al.}(2020)Zorrilla~Matilla, Sharma, Hsu,
  \& Haiman}]{Matilla}
Zorrilla~Matilla, J.~M., Sharma, M., Hsu, D., \& Haiman, Z. 2020,
  \bibinfo{title}{{Interpreting deep learning models for weak lensing},} Phys.
  Rev. D, 102, 123506, \dodoi{10.1103/PhysRevD.102.123506}

\end{thebibliography}
\bibliographystyle{aasjournalv7}



\end{document}